\documentclass[aps,prd]{revtex4}

\usepackage{amsmath}
\usepackage{amssymb}
\usepackage{bm}% bold math
\usepackage{graphicx}
\usepackage{multirow}

\usepackage{textcomp}

\allowdisplaybreaks[1]

\begin{document}
%\preprint{  }

\title{Dispersive determination of the fourth generation lepton masses}
\author{Hsiang-nan Li\footnote{Corresponding author, E-mail: hnli@phys.sinica.edu.tw}}
\affiliation{Institute of Physics, Academia Sinica,
Taipei, Taiwan 115, Republic of China}

\date{\today}

\begin{abstract}

We continue our previous determination of the masses of the sequential fourth generation 
quarks in an extension of the Standard Model, and predict the mass $m_4$ ($m_L$) of the 
fourth generation neutrino $\nu_4$ (charged lepton $L$) by solving the dispersion 
relations associated with heavy fermion decays. The results $m_4\approx 170$ GeV and
$m_L\approx 270$ GeV are extracted from the dispersive analyses of the $t\to d e^+\nu_4$ 
and $L^-\to \nu_1 \bar t d$ decay widths, respectively, where $t$ ($d$, $e^+$, $\nu_1$) 
denotes a top quark (down quark, positron, light neutrino). The predictions are 
cross-checked by examining the $L^-\to \nu_4 \bar u d$ decay, $\bar u$ being an anti-up 
quark. It is shown that the fourth generation leptons with the above masses survive the 
current experimental bounds from Higgs boson decays into photon pairs and from the oblique 
parameters. We also revisit how the existence of the fourth generation leptons impacts the
dispersive constraints on the neutrino masses and the Pontecorvo–Maki–Nakagawa–Sakata 
(PMNS) matrix elements. It is found that the unitarity of the $3\times 3$ PMNS matrix holds 
well up to corrections of $O(m_\nu^2/m_W^2)$, $m_\nu$ ($m_W$) being a light neutrino 
(the $W$ boson) mass, whose mixing angles and $CP$ phase prefer the values 
$\theta_{12}\approx 34^\circ$, $\theta_{23}\approx 47^\circ$, $\theta_{13}\approx 5^\circ$ 
and $\delta\approx 200^\circ$ in the normal-ordering scenario for neutrino masses. 

%On the contrary, the unitarity of the $3\times 3$ Cabibbo-Kobayashi-Maskawa matrix is 
%violated at the level $m_b^2/m_W^2\sim 10^{-3}$, $m_b$ being the $b$ quark mass, as the 
%same formalism is applied to the quark mixing.

\end{abstract}

\maketitle

%--------+---------+---------+---------+---------+---------+---------+---------+
\section{INTRODUCTION}

We have performed dispersive analyses on the flavor structure of the Standard Model (SM)
in a series of publications recently \cite{Li:2023dqi,Li:2023yay,Li:2023ncg,Li:2024awx}. 
Sufficient clues for understanding the mass hierarchy and the distinct mixing patters of 
quarks and leptons have been accumulated, which suggest that the scalar sector could be 
stringently constrained by the internal consistency of SM dynamics. We were then motivated 
to explore the sequential fourth generation model as a natural and simple extension of the 
SM, for which no additional free parameters need to be introduced. The masses of the 
fourth generation quarks $b'$ and $t'$ were determined by studying neutral quark state 
mixing through box diagrams, i.e., the $Q\bar q$-$\bar Q q$ mixing with $Q$ ($q$) being a 
heavy (light) quark \cite{Li:2023fim}. The idea is to treat the dispersion relation obeyed 
by a mixing observable as an inverse problem with the input from the perturbative evaluation 
of the box diagrams \cite{Li:2020xrz,Li:2022jxc,Xiong:2022uwj}. It was demonstrated that a 
solution to the dispersion relation demands a specific mass of the heavy quark 
\cite{Li:2023fim}; the common mass $m_{b'}\approx 2.7$ TeV ($m_{t'}\approx 200$ TeV) was  
obtained from the dispersive analyses of the box diagrams involving the intermediate quark
channels $ut$ and $ct$ ($db'$, $sb'$ and $bb'$).

For an extension of the SM to be anomaly free, the existence of the fourth generation quarks 
requires that of the fourth generation leptons. The intensive practice of our formalism 
guarantees that the masses of the fourth generation leptons can also be predicted by 
investigating appropriate observables, such as heavy fermion decays. The $Z$ boson decay data 
\cite{PDG} have indicated that the fourth generation neutrino $\nu_4$ and charged lepton 
$L$ are heavier than half of the $Z$ boson mass. Hence, we examine the dispersion 
relation for the semileptonic top quark decay $t\to d e^+\nu_4$ in the framework similar 
to the one in \cite{Li:2023dqi}, $d$ ($e^+$) being a down quark (positron), and show that 
the reproduction of the top quark mass $m_t\approx 173$ GeV leads to the $\nu_4$ mass 
$m_4\approx 170$ GeV. The study of the hadronic decay $L^-\to \nu_i \bar t d$, where 
$\nu_i$, $i=1$, 2 or 3, represents a light neutrino, yields the $L$ mass $m_L\approx 270$ 
GeV, as $m_t\approx 173$ GeV is input into the corresponding dispersion relation. We then 
substantiate the above heavy lepton masses by checking the dispersion 
relation for the hadronic decay $L^-\to \nu_4 \bar u d$ with an anti-up quark $\bar u$. A 
positron, light neutrinos, and light quarks in final states are all treated as massless 
particles \cite{Li:2023dqi} for simplicity. The simultaneous satisfaction of the dispersion 
relations for the three different processes by our predictions should not be a coincidence.

The merits of the sequential fourth-generation model have been summarized in 
\cite{Li:2023fim}; it provides a dynamical mechanism for electroweak symmetry breaking by 
means of heavy fermion condensates \cite{HBH,Mimura:2012vw}, realizes electroweak 
baryogenesis through the first-order phase transition \cite{HOS}, and offers a viable source 
of $CP$ violation for the baryon asymmetry of the Universe \cite{Hou:2008xd}. It was pointed 
out that the fourth generation quarks $b'$ and $t'$ with the masses above a TeV scale form 
bound states in a Yukawa potential \cite{Hung:2009hy,Enkhbat:2011vp}. The contributions from 
the $\bar b'b'$ scalars to the Higgs boson production via gluon fusion and to the Higgs decay 
into a photon pair, $H\to \gamma\gamma$, were estimated to be of $O(10^{-3})$ and $O(10^{-2})$ 
of the top quark one \cite{Li:2023fim}, respectively. These estimates illustrated why these 
superheavy quarks bypass the experimental constraints from Higgs boson production and 
decay \cite{Chen:2012wz}. The fourth generation leptons, with the masses of the electroweak 
scale, are too light to form bound states. We thus calculate the contribution from the charged 
lepton of the mass 270 GeV to the $H\to \gamma\gamma$ decay width $\Gamma_{\gamma\gamma}$, and 
evince that its effect is within the current uncertainties of $\Gamma_{\gamma\gamma}$ 
measurements. The impact on the oblique parameters from the fourth generation quarks and 
leptons, inspected based on the formulas supplied in \cite{He:2001tp}, is also 
allowed by the experimental errors of the oblique parameters. The search for heavy neutral 
leptons at colliders within the mass range from GeV to the electroweak scale has been updated 
in \cite{Marcano:2024bjs}. 

%\cite{Carbajal:2022zlp}, Indirect search of heavy neutral leptons using the DUNE near detector

At last, we revisit how the existence of the fourth generation leptons modifies the 
constraints on the neutrino masses and the Pontecorvo–Maki–Nakagawa–Sakata (PMNS) matrix 
elements. These constraints were derived for the case with three generations of neutrinos 
in \cite{Li:2023ncg}, which favor the neutrino masses in the normal ordering (NO), instead of 
the inverted ordering (IO), in view of the consistency with the observed PMNS matrix elements
\cite{deSalas:2017kay,Capozzi:2018ubv}. 
We repeat the analysis with four generations of neutrinos here, and find that the unitarity 
of the $3\times 3$ PMNS matrix holds quite well. Namely, the fourth generation leptons barely 
mix with the three light generations. The theoretical preference on the NO scenario is 
confirmed, and the aforementioned constraints are better respected. Explicitly speaking, the 
known mass squared differences among the neutrinos are linked to the mixing angles 
$\theta_{12}\approx 34^\circ$, $\theta_{23}\approx 47^\circ$ and $\theta_{13}\approx 5^\circ$, 
and the $CP$ phase $\delta\approx 200^\circ$, close to the measured values for the NO scenario. 
In this sense, the data for the PMNS mixing angles have revealed the possible existence 
of the fourth generation leptons. As the same formalism is applied to the quark mixing, 
the unitarity of the $3\times 3$ Cabibbo-Kobayashi-Maskawa matrix is expected to be 
violated at the level of $m_b^2/m_W^2\sim 10^{-3}$, $m_b$ ($m_W$) being the $b$ quark ($W$ 
boson) mass. 

%We mention that the existence of the fourth generation neutrino may help resolve 
%the so-called Hubble tension.

The rest of the paper is organized as follows. We briefly recollect the construction of a 
dispersion relation based on analyticity of a physical observable and its solution 
in Sec.~II. The masses of the fourth generation leptons are 
extracted from the dispersion relations for the $t\to d e^+\nu_4$ and $L^-\to \nu_1 \bar t d$ 
decays, and cross-checked by means of the $L^-\to \nu_4 \bar u d$ decay in 
Sec.~III. The experimental bounds from Higgs boson decays into photon pairs and from the 
oblique parameters on the sequential fourth generation model are scrutinized in Sec.~IV. 
We discuss the dispersive constraints on the neutrino masses and the PMNS matrix elements 
in the presence of the fourth generation leptons in Sec.~V. The mixing 
angles and the $CP$ phase involved in the $3\times 3$ PMNS matrix are solved for and confronted 
with the data. Section~VI contains the conclusion; our observations encourage the  
search for the fourth generation leptons at the (high-luminosity) large hadron collider or a 
muon collider.

\section{DISPERSION RELATION AND ITS SOLUTION}

\begin{figure}
\begin{center}
\includegraphics[scale=0.7]{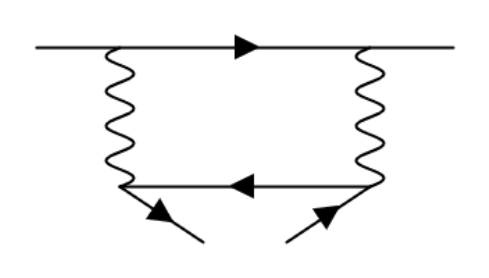}

\caption{\label{fig1}
Box diagram for a heavy fermion decay, where the wavy lines represent $W$ bosons. 
%(b) Contour for Eq.~(\ref{con}), where the thick lines represent the branch cuts.
}
\end{center}
\end{figure}

Consider the analytical amplitude $\Pi(m_Q^2)$ associated with the box diagram in 
Fig.~\ref{fig1}, whose imaginary part gives the inclusive decay width $\Gamma(m_Q)$ 
of a heavy fermion $Q$ of the mass $m_Q$. The diagram describes the semileptonic decay 
$t\to dW^+\to d e^+\nu_4$ or the hadronic decay $L^-\to\nu_1W^-\to \nu_1 \bar t d$ to be 
explored in the next section. Here the first generation neutrino $\nu_1$ is representative 
and can be replaced by $\nu_2$ or $\nu_3$. Because a top quark is heavier than a $W$ boson, 
the latter cannot be integrated out to get the effective weak Hamiltonian for the above 
decays. The operator definition of the amplitude $\Pi(m_Q^2)$ is thus lengthier 
than in Ref.~\cite{Li:2023dqi}, where decays of lighter fermions were studied. For example,
the amplitude related to the semileptonic decay $t\to d e^+\nu_4$ is written as
\begin{eqnarray}
\Pi(m_t^2)\propto\frac{1}{2m_t}\int d^4x H^{\mu\nu}(x)
\langle t|J_\mu^\dagger(x)J_\nu(0)|t\rangle,\label{1}
\end{eqnarray}
in which the function 
\begin{eqnarray}
H^{\mu\nu}(x)=\int \frac{d^4k_1}{(2\pi)^4}\frac{d^4k_2}{(2\pi)^4}
\exp[-ix\cdot (k_1+k_2)]\frac{k_1^\mu k_2^\nu+k_1^\nu k_2^\mu-g^{\mu\nu}k_1\cdot k_2}
{k_1^2(k_2^2-m_4^2)[(k_1+k_2)^2-m_W^2]^2},
\end{eqnarray}
collects the contribution from the leptonic part, and $J_\mu=\bar d\gamma_\mu P_Lt$ with 
the projector $P_L=(1-\gamma_5)/2$ is the $V-A$ current. The overall coefficient, irrelevant 
to the discussion, has been omitted. The construction of Eq.~(\ref{1}) follows the 
one for the Drell-Yan cross section in \cite{Manohar:2012jr}.

We have the identity from the contour integration 
for $\Pi(m_Q^2)$ \cite{Li:2022jxc}, 
\begin{eqnarray}
\frac{1}{2\pi i}\oint \frac{\Pi(m^2)}{m^2-m_Q^2}dm^2=0,\label{con}
\end{eqnarray}
whose contour consists of two pieces of horizontal paths above and below the branch cut 
along the positive real axis, a small circle around the pole $m^2=m_Q^2$ located on the 
positive real axis and a circle $C_R$ of the large radius $R$. The above integral vanishes, 
for the contour encloses unphysical regions without poles.
Equation~(\ref{con}) is rewritten as
\begin{eqnarray}
{\rm Re}\Pi(m_Q^2)=\frac{1}{\pi}
\int_{m_F^2}^{R^2} \frac{{\rm Im}\Pi(m^2)}{m^2-m_Q^2}dm^2
%-\frac{1}{\pi}\int^{-m_F}_{-R} \frac{{\rm Im}\Pi(m)}{m-m_Q}dm
+\frac{1}{2\pi i}\int_{C_R} \frac{\Pi^{\rm p}(m^2)}{m^2-m_Q^2}dm^2,\label{ij}
\end{eqnarray}
where the hadronic threshold $m_F$ sums the masses in the lightest final state. The 
contribution along the small clockwise circle gives the real part ${\rm Re}\Pi(m_Q^2)$ on 
the left-hand side. The contribution from the horizontal paths along the positive real 
axis leads to the dispersive integral of the imaginary part ${\rm Im}\Pi(m^2)$. It 
contains nonperturbative dynamics associated with light hadronic final states 
near the threshold, where the decay becomes exclusive. The integrand $\Pi(m^2)$, taking 
values along the large counterclockwise 
circle $C_R$, has been reliably approximated by $\Pi^{\rm p}(m^2)$ from the perturbative 
evaluation of Fig.~\ref{fig1}. The perturbative result $\Pi^{\rm p}(m_Q^2)$ from 
Fig.~\ref{fig1} certainly obeys a similar dispersion relation because of the analyticity,
\begin{eqnarray}
{\rm Re}\Pi^{\rm p}(m_Q^2)=\frac{1}{\pi}
\int_{m_f^2}^{R^2} \frac{{\rm Im}\Pi^{\rm p}(m^2)}{m^2-m_Q^2}dm^2
%-\frac{1}{\pi}\int^{-m_f}_{-R} \frac{{\rm Im}\Pi^{\rm p}(m)}{m-m_Q}dm
+\frac{1}{2\pi i}\int_{C_R} \frac{\Pi^{\rm p}(m^2)}{m^2-m_Q^2}dm^2,\label{ope}
\end{eqnarray}
with the quark-level threshold $m_f$ in the dispersive 
integral, above which the imaginary part ${\rm Im}\Pi^{\rm p}(m^2)$ is defined. 

%\begin{eqnarray}
%\int_{m_F}^R\frac{{\rm Im}\Pi(m)}{m-m_Q}dm-
%\int^{-m_F}_{-R}\frac{{\rm Im}\Pi(m)}{m-m_Q}dm=
%\int_{m_f}^R\frac{{\rm Im}\Pi^{\rm p}(m)}{m-m_Q}dm-
%\int^{-m_f}_{-R}\frac{{\rm Im}\Pi^{\rm p}(m)}{m-m_Q}dm.\label{ij2}
%\end{eqnarray} 
%The contributions from the large circle $C_R$ on the two sides have canceled each other. 
%The above formula indicates that the amplitude ${\rm Im}\Pi(m)$ (${\rm Im}\Pi^{\rm p}(m)$) 
%needs to be even functions of $m$, and defined as a product of $\Gamma(m)$ 
%($\Gamma^{\rm p}(m)$ with an odd function of $m$ \cite{Li:2023dqi}, as shown in the next
%section. We then apply the variable change $m\to -m$ to the second integrals on both sides 
%of Eq.~(\ref{ij1}), arriving at the standard form in the integration variable $m^2$,

We equate ${\rm Re}\Pi(m_Q^2)$ and ${\rm Re}\Pi^{\rm p}(m_Q^2)$, i.e., Eqs.~(\ref{ij}) and 
(\ref{ope}) at large enough $m_Q\gg m_F$, where a perturbation theory is trustworthy, 
arriving at 
\begin{eqnarray}
\int_{m_F^2}^{R^2}\frac{{\rm Im}\Pi(m^2)}{m^2-m_Q^2}dm^2=
\int_{m_f^2}^{R^2}\frac{{\rm Im}\Pi^{\rm p}(m^2)}{m^2-m_Q^2}dm^2.\label{ij1}
\end{eqnarray} 
The pieces from the large circle $C_R$ on the two sides of the equation have canceled. The 
distinction between the lower bounds $m_F^2$ and $m_f^2$ forces ${\rm Im}\Pi(m^2)$ to deviate 
from ${\rm Im}\Pi^{\rm p}(m^2)$ under the above dispersion
relation. Moving the integrand on the right-hand side to the left-hand side, and regarding 
it as a subtraction term, we attain
\begin{eqnarray}
\int_{m_f^2}^\infty\frac{\Delta\rho(m^2)}{m_Q^2-m^2}dm^2=0.\label{ge}
\end{eqnarray}
The subtracted unknown function 
$\Delta\rho(m^2)\equiv{\rm Im}\Pi(m^2)-{\rm Im}\Pi^{\rm p}(m^2)$ is set to 
$-{\rm Im}\Pi^{\rm p}(m^2)$ in the interval $(m_f^2,m_F^2)$ of the variable $m^2$ as an 
initial condition, and approaches zero at large $m^2$, owing to 
${\rm Im}\Pi(m^2)\to{\rm Im}\Pi^{\rm p}(m^2)$ in this limit. Accordingly, the radius $R$ of 
the large circle has been extended to infinity. Since Eq.~(\ref{ge}) holds for any large 
$m_Q$, it imposes a strict connection between the nonperturbative behavior of 
${\rm Im}\Pi(m^2)$ near the threshold and the perturbative ${\rm Im}\Pi^{\rm p}(m^2)$. 

The variable changes $m_Q^2-m_f^2=u\Lambda$ and $m^2-m_f^2=v\Lambda$, $\Lambda$ being an 
arbitrary scale, turn Eq.~(\ref{ge}) into
\begin{eqnarray}
\int_{0}^\infty dv\frac{\Delta\rho(v)}{u-v}=0.\label{i2}
\end{eqnarray}
The arbitrariness of the radius $R$ in Eq.~(\ref{ij1}) has been switched to that of
the scale $\Lambda$ in some sense.
We request that $\Delta\rho(v)$ diminishes quickly (to be precise, exponentially) at large 
$v$; namely, ${\rm Im}\Pi(m^2)$ approaches ${\rm Im}\Pi^{\rm p}(m^2)$ quickly at large $m^2$. 
The fast diminishing of $\Delta\rho(v)$ at high $v$ is one of the boundary conditions for 
solving the above integral equation (another is specified in the low end of $v$, i.e., in the 
interval $(0,(m_F^2-m_f^2)/\Lambda)$ below). The major contribution to Eq.~(\ref{i2}) then 
arises from the region with finite $v$, such that Eq.~(\ref{i2}) can be expanded into a power 
series in $1/u$ for sufficiently large $u$ by inserting $1/(u-v)=\sum_{k=1}^\infty v^{k-1}/u^k$. 
Equation~(\ref{i2}) thus demands a vanishing coefficient for every power of $1/u$ on account 
of the arbitrariness of $u$, from which a solution can be built up. This method for solving 
an integral equation has a solid ground in mathematics, as sketched in \cite{CWG}. The setup 
has been applied to the explanation of light fermion and electroweak boson masses successfully 
\cite{Li:2023dqi,Li:2023yay}. 

We start with the case with $N$ vanishing coefficients,
\begin{eqnarray}
\int_{0}^\infty dvv^{k-1}\Delta\rho(v)=0,\;\;\;\;k=1,2,3\cdots,N,\label{i3}
\end{eqnarray}
where $N$ will be sent to infinity eventually. The first $N$ generalized Laguerre  
polynomials $L_{0}^{(\alpha)}(v)$, $L_{1}^{(\alpha)}(v)$, $\cdots$, $L_{N-1}^{(\alpha)}(v)$ 
are composed of the terms $1$, $v$, $\cdots$, $v^{N-1}$ appearing in the above expressions. 
Therefore, Eq.~(\ref{i3}) implies an expansion of $\Delta\rho(v)$ in $L_k^{(\alpha)}(v)$ 
with degrees $k$ not lower than $N$,
\begin{eqnarray}
\Delta \rho(v)=\sum_{k=N}^{N'} a_kv^{\alpha} e^{-v}
L_{k}^{(\alpha)}(v),\;\;\;\;N'>N,\label{d0}
\end{eqnarray}
attributed to the orthogonality of the polynomials, in which $a_k$ represent a set of unknown 
coefficients. The index $\alpha$ characterizes the behavior of $\Delta \rho(v)$ around the 
boundary $v\sim 0$. The highest degree $N'$ can be fixed in principle by the initial 
condition $-{\rm Im}\Pi^{\rm p}(v)$ of $\Delta \rho(v)$ in the interval 
$(0,(m_F^2-m_f^2)/\Lambda)$ of $v$. Because $-{\rm Im}\Pi^{\rm p}(v)$ is a 
smooth function, $N'$ needs not be infinite. As proved in \cite{Xiong:2022uwj}, the integral 
equation in Eq.~(\ref{i2}) with specified boundary conditions, i.e., the so-called 
Fredholm equation of the first kind, has a unique solution, which describes the 
properties of $\Delta\rho(v)$ in the whole range of $v$.

%. Once we fix $N'$ in the solution 
%by the above initial condition, it is the unique solution faithfully

A generalized Laguerre polynomial takes the asymptotic form for large $k$,
$L_k^{(\alpha)}(v)\approx k^{\alpha/2}v^{-\alpha/2}e^{v/2}J_\alpha(2\sqrt{kv})$ \cite{BBC},
up to corrections of $O(1/\sqrt{k})$ with $J_\alpha$ being a Bessel function. 
Equation~(\ref{d0}) becomes
\begin{eqnarray}
\Delta \rho(m^2)
\approx\sum_{k=N}^{N'} a_k\sqrt{\frac{k(m^2-m_f^2)}{\Lambda}}^{\alpha} 
e^{-(m^2-m_f^2)/(2\Lambda)}J_{\alpha}\left(2\sqrt{\frac{k(m^2-m_f^2)}{\Lambda}}\right),
\label{d1}
\end{eqnarray}
where the argument $m^2=v\Lambda+m_f^2$ has been recovered. Defining the scaling variable 
$\omega\equiv\sqrt{N/\Lambda}$, we have the approximation 
$N'/\Lambda=\omega^2[1+(N'-N)/N]\approx \omega^2$ for finite $N'-N$. The common Bessel 
functions $J_{\alpha}\left(2\sqrt{k(m^2-m_f^2)/\Lambda}\right)\approx J_{\alpha}
\left(2\omega \sqrt{m^2-m_f^2}\right)$ for $k=N,N+1,\cdots,N'$ can then be factored out, 
such that the unknown coefficients are summed into a single parameter $a=\sum_{k=N}^{N'}a_k$. 
It has been verified \cite{Li:2023dqi,Li:2023yay,Li:2023fim} that a single ($N'=N$) Bessel 
function $J_{\alpha}$ can fit the initial condition in the interval $(m_f^2,m_F^2)$ well.
The arbitrariness of $\Lambda$, tracing back to that of the radius $R$, goes into the variable
$\omega$. We are permitted to treat $\omega$ as a finite quantity, though both $N$ and 
$\Lambda$ are very large. The exponential suppression factor 
$e^{-(m^2-m_f^2)/(2\Lambda)}=e^{-\omega^2 (m^2-m_f^2)/(2N)}$ in Eq.~(\ref{d1}) is further 
replaced by unity for finite $\omega$ and high $N$. 

We establish the solution to the integral equation in Eq.~(\ref{ge}),
\begin{eqnarray}
\Delta \rho(m_Q^2)\approx a\left(\omega\sqrt{m_Q^2-m_f^2}\right)^{\alpha} 
J_{\alpha}\left(2\omega \sqrt{m_Q^2-m_f^2}\right),\label{so3}
\end{eqnarray}
whose parameters can be determined by the boundary conditions. First, Eq.~(\ref{so3}) scales 
in the threshold region with $m_Q\sim m_f$ like $\Delta \rho(m_Q^2)\propto (m_Q^2-m_f^2)^{\alpha}$ 
according to the approximation $J_{\alpha}(z)\propto z^\alpha$ in the limit $z\to 0$. Contrasting 
this scaling law near the threshold with that of the initial condition 
$-{\rm Im}\Pi^{\rm p}(m_Q^2)$, we read off the index $\alpha$.
Another condition $\Delta \rho(m_F^2)=-{\rm Im}\Pi^{\rm p}(m_F^2)$ at $m_Q=m_F$ sets 
the overall coefficient 
\begin{eqnarray}
a=-{\rm Im}\Pi^{\rm p}(m_F^2)\left[\left(\omega\sqrt{m_F^2-m_f^2}\right)^{\alpha}
J_\alpha\left(2\omega\sqrt{m_F^2-m_f^2}\right)\right]^{-1}.\label{dc4}
\end{eqnarray}

%It has been known that an ill-posed integral equation is difficult to solve.
%The above inverse matrix method corresponds to the special case that the kernel is degenerate
%$k(x,t)=\sum_{j=1}^n X_j(x)T_j(t)$. We can take $T_j$ to be linearly independent.
%Then the integral equation has a solution if and only if the input can be cast into
%$g\in {\rm span}(X_1,X_2,\cdots,X_n)$. In this case solving the integral equation 
%reduces to the problem of solving a finite system of linear equations.

A solution for the physical decay width must be insensitive to the arbitrary variable $\omega$.
To realize this insensitivity, we make a Taylor expansion of $\Delta \rho(m_Q^2)$ 
\cite{Li:2023yay},
\begin{eqnarray}
\Delta \rho(m_Q^2)=\Delta \rho(m_Q^2)|_{\omega=\bar\omega}+
\frac{d\Delta \rho(m_Q^2)}{d\omega}\Big|_{\omega=\bar\omega}(\omega-\bar\omega)+
\frac{1}{2}\frac{d^2\Delta \rho(m_Q^2)}{d\omega^2}\Big|_{\omega=\bar\omega}
(\omega-\bar\omega)^2+\cdots,\label{ta}
\end{eqnarray}
where the constant $\bar\omega$, together with the index $\alpha$ and the coefficient 
$a$, are fixed through the fit of the first term $\Delta \rho(m_Q^2)|_{\omega=\bar\omega}$ 
to the initial condition in the interval $(m_f^2,m_F^2)$ of $m_Q^2$. The insensitivity to 
the scaling variable $\omega$ is achieved by vanishing the first derivative in Eq.~(\ref{ta}),
\begin{eqnarray}
\frac{d\Delta \rho(m_Q^2)}{d\omega}\Big|_{\omega=\bar\omega}=0,\label{dd1}
\end{eqnarray}
from which roots of $m_Q$ are solved. Moreover, the second derivative 
$d^2\Delta \rho(m_Q^2)/d\omega^2|_{\omega=\bar\omega}$ needs to be minimal to maximize the 
stability window around $\bar\omega$, in which $\Delta\rho(m_Q^2)$ is almost independent of 
the variation of $\omega$. As a consequence, only when $m_Q$ takes a specific 
value, can the above requirements be met. Once a solution is constructed, the degree $N$ for the 
polynomial expansion in Eq.~(\ref{d0}) and the scale $\Lambda$ can approach infinity by 
maintaining $\omega=\sqrt{N/\Lambda}$ within the stability window. Then all the arguments 
based on the large $N$ assumption, including the neglect of the exponential factor 
$e^{-(m^2-m_f^2)/(2\Lambda)}$ in Eq.~(\ref{d1}), are justified.

\section{MASSES OF THE FOURTH GENERATION LEPTONS}

We apply the formalism outlined in the previous section to the analyses of the 
$t\to d e^+\nu_4$, $L^-\to \nu_1 \bar t d$ and $L^-\to \nu_4 \bar u d$ decay widths, and 
determine the masses $m_4$ and $m_L$ of the fourth generation leptons.

\subsection{$m_4$ from the $t\to d e^+\nu_4$ Decay}

Since a CKM matrix element can vary independently in a mathematical point of view, heavy 
quark decays with different CKM matrix elements, i.e., into distinct final states, 
follow separate dispersion relations. It is thus legitimate to acquire a fermion mass by 
analyzing appropriate modes. The application to the semileptonic 
decay $b\to u\tau^- \bar\nu_\tau$, $\tau$ ($\nu_\tau$) being a $\tau$ lepton (neutrino), 
has generated the expected $b$ quark mass $m_b\approx 4$ GeV, given the $\tau$ lepton mass 
$m_\tau\approx 2$ GeV \cite{Li:2023dqi}. It was pointed out in the Introduction that the 
fourth generation leptons are heavier than half of the $Z$ boson mass. We attempt the scenario 
that the fourth generation neutrino $\nu_4$ is lighter than the charged lepton $L$ and than 
a top quark, and constrain its mass $m_4$ using the $t\to d e^+\nu_4$ decay width. The mass 
$m_4$ is identified as the one, which reproduces the top quark mass $m_t\approx 173$ GeV 
\cite{PDG} from the corresponding dispersion relation. This strategy is analogous to the one 
adopted for the $b\to u\tau^- \bar\nu_\tau$ decay \cite{Li:2023dqi}. The 
ordering of $m_4$, $m_t$ and $m_L$ will be confirmed at the end of this section.

The width of the semileptonic decay $t\to bW^+\to b \ell^+\nu_\ell$ has been obtained 
in \cite{Ali:2010auk} up to the $O(\alpha_s)$ correction with the strong coupling 
constant $\alpha_s$ \cite{Jezabek:1988ja}. For related higher-order QCD calculations
of the $t\to bW$ decay width, refer to the recent publication \cite{Yan:2024hbz}. We first 
deduce the tree-level expression for the $t\to de^+\nu_4$ decay width from the formulas in 
\cite{Chang:1998ze}, and then incorporate the $O(\alpha_s)$ correction \cite{Jezabek:1988ja}, 
where the $m_b$-dependent terms are removed to match the considered mode with a 
massless down quark. Define the energy fraction $x_e= 2E_e/m_t$ of the positron $e^+$ and the 
dimensionless variable $y_W= q^2/m_t^2$, $q^2$ being the invariant mass squared of the $W$ 
boson, i.e., the $e^+\nu_4$ pair. It is trivial to specify their kinematic ranges 
\begin{eqnarray}
0\le &x_e& \le 1-\eta,\nonumber\\
\eta\left(1+\frac{x_e}{1-x_e}\right)\le &y_W& \le x_e+\eta,\label{ran}
\end{eqnarray}
with the notation $\eta= m_4^2/m_t^2$. The involved hard kernel is proportional to 
\cite{Chang:1998ze}
\begin{eqnarray}
(p_t\cdot p_e)(p_d\cdot p_{\nu_4})\propto x_e(1-\eta-x_e),\label{h1}
\end{eqnarray}
where $p_t$ ($p_e$, $p_d$, $p_{\nu_4}$) is the momentum of the top quark $t$ (the 
positron $e^+$, the down quark $d$, the heavy neutrino $\nu_4$).

The $t\to de^+\nu_4$ decay width from the perturbative evaluation is then written as
\begin{eqnarray}
\Gamma^{\rm p}(m_t)\propto G_F^2m_t^5(1-\eta)^4\int_0^1dxx
\int_0^1\frac{m_W^4dy}{[m_W^2-m_t^2y_W(x,y)]^2+\Gamma_W^4}
\left[(1-\eta)x(1-x)-\frac{2\alpha_s(m_t)}{3\pi}F_t(x)\right],\label{ms}
\end{eqnarray} 
with $m_W$ ($\Gamma_W$) being the mass (width) of a $W$ boson. The overall constant coefficient, 
including the Fermi constant $G_F$, the CKM matrix element $V_{td}$ and the PMNS matrix element 
$U_{e4}$, is not crucial, in that it is canceled from the two sides of Eq.~(\ref{ij1}). The range 
of the integration variable $x$ has been scaled into $[0,1]$ via $x_e=(1-\eta)x$. The effective 
weak Hamiltonian with the $W$ boson mass being integrated out was employed in the studies 
of bottom or charm quark decays \cite{Li:2023dqi}. Here we need to retain a $W$ boson propagator, 
which gives rise to the denominator in Eq.~(\ref{ms}). The $W$ boson invariant mass squared is
expressed, in terms of $y_W$, as
\begin{eqnarray}
y_W(x,y)=(1-\eta)^2x\left(y + \frac{\eta}{1-\eta}\right) + \eta,\label{yw}
\end{eqnarray}
which can be regarded as a variable change from $y_W$ to $y$. The argument of the running 
coupling constant has been set to $\mu=m_t$. The function collecting the QCD correction 
reads \cite{Jezabek:1988ja}
\begin{eqnarray}
F_t(x,y)\approx (1-\eta)x(1-x)\ln^2[(1-\eta)(1-x)]+\frac{5}{2}(1-x)\ln[(1-\eta)(1-x)],\label{ft1}
\end{eqnarray}
where only the dominant threshold logarithmic terms are taken into account. 
It has been verified that keeping the other $O(\alpha_s)$ pieces changes our prediction 
for $m_4$ by less than 1\%.

%F_t
%& &+\frac{x}{2}\ln(1-y)\{9-4(1-\eta_t)x-2y[1+(1-\eta_t)x]-(1-\eta_t) xy^2\},
%F_W(x,y)&=&2x(1-\eta-x}\left[\frac{\pi^2}{6}+Li_2(x)+Li_2(y/x)
%+\frac{1}{2}\ln^2\frac{1-y/x}{1-\eta-x}\right]\nonumber\\
%& &+x\left[\frac{\pi^2}{6}+Li_2(y)-Li_2(x)-Li_2(y/x)\right]\nonumber\\
%& &+\frac{1}{2}\ln(1-y)[y^2+2y(1+x)-(3+2x)]\nonumber\\
%& &+\frac{1}{2}\ln(1-y/x)[x(9-4x)-2y(1+x)-y^2]\nonumber\\
%& &+\frac{5}{2}(1-x)\ln(1-x)+\frac{1}{2}y(1-x)\left(\frac{y}{x}+4\right).

We treat the top quark mass $m_t$ as a variable $m_Q$ and set the perturbative threshold
$m_f$ to $m_4$ in this case (both $d$ and $e^+$ are massless) as implementing the formalism. 
Equation~(\ref{ms}) behaves, around $m_Q\sim m_f$, like 
\begin{eqnarray}
\Gamma^{\rm p}(m_Q)\propto \frac{(m_Q^2-m_f^2)^4}{m_Q^3},\label{asy}
\end{eqnarray}
which motivates the choice of the integrands in Eq.~(\ref{ij1}) \cite{Li:2023yay},
\begin{eqnarray}
{\rm Im}\Pi^{({\rm p})}(m^2)&=&\frac{m^3\Gamma^{({\rm p})}(m)}{(m^2-m_f^2)^2}.\label{mi}
\end{eqnarray}
The above expression with a power of $m$ in the numerator suppresses potential residues in 
the low $m$ region, including that from the pole at $m^2=m_f^2$, relative to $m^2=m_Q^2$ 
at large $m_Q$. The denominator alleviates the enhancement caused by the modified numerator 
at large $m$. The integrands in Eq.~(\ref{mi}), as even functions of the variable $m$, 
approves the derivation of the dispersion relation starting with Eq.~(\ref{con}). 
Equations~(\ref{asy}) and (\ref{mi}) then prescribe the boundary condition of the subtracted 
unknown function in the limit $m_Q\to m_f$,
\begin{eqnarray}
\Delta\rho(m_Q^2)\propto (m_Q^2-m_f^2)^2.\label{p2}
\end{eqnarray}
Comparing the scaling law of Eq.~(\ref{so3}), $\Delta \rho(m_Q^2)\propto (m_Q^2-m_f^2)^{\alpha}$,
with Eq.~(\ref{p2}), we designate the index $\alpha=2$; the 
integrands in Eq.~(\ref{mi}) make the input in Eq.~(\ref{p2}) proportional to a simple 
power of $m_Q^2-m_f^2$, such that the index $\alpha$ can be read off straightforwardly.

The strong coupling constant is given by
\begin{eqnarray}
\alpha_s(\mu)=\frac{4\pi}{\beta_0\ln(\mu^2/\Lambda_{\rm QCD}^2)},\label{rs}
\end{eqnarray}
with the coefficient $\beta_0 = 11 - 2n_f/3$. We take the QCD scale parameter 
$\Lambda_{\rm QCD}=0.207$ GeV for the number of active quarks $n_f=5$ 
\cite{Bruno:2016zeo}. The interval, in which the initial condition is defined, spans from 
the lower bound $m_f=m_4$ to the upper bound $m_F=m_4+m_{\pi^0}$, because a $\pi^0$ meson 
is the lightest one among the hadronic bound states formed by a down quark. Inputting 
$\Gamma_W = 2.085$ GeV, $m_W = 80.4$ GeV, $m_{\pi^0}=0.135$ GeV \cite{PDG} and 
$m_4= 170$ GeV, we get the parameter $\bar\omega=0.1001$ GeV$^{-1}$ from the best 
fit of $\Delta\rho(m_Q^2)|_{\omega=\bar\omega}$ in Eq.~(\ref{ta}), with $\Delta\rho(m_Q^2)$
in Eq.~(\ref{so3}), to $-{\rm Im}\Pi^{\rm p}(m_Q^2)$ in Eq.~(\ref{mi}), with 
$\Gamma^{{\rm p}}(m_Q)$ in Eq.~(\ref{ms}), in the interval $(m_f^2,m_F^2)$.
As seen shortly, this value of $m_4$, above half of a $Z$ boson mass, 
yields the top quark mass $m_t\approx 173$ GeV. The fit quality is as good as that exhibited
in \cite{Li:2023dqi,Li:2023fim}, and will not be displayed explicitly. The subtracted 
unknown function $\Delta\rho(m_Q^2)$ shows an oscillatory behavior in $m_Q$, also 
similar to what was observed in \cite{Li:2023dqi,Li:2023fim}.

\begin{figure}
\begin{center}
\includegraphics[scale=0.2]{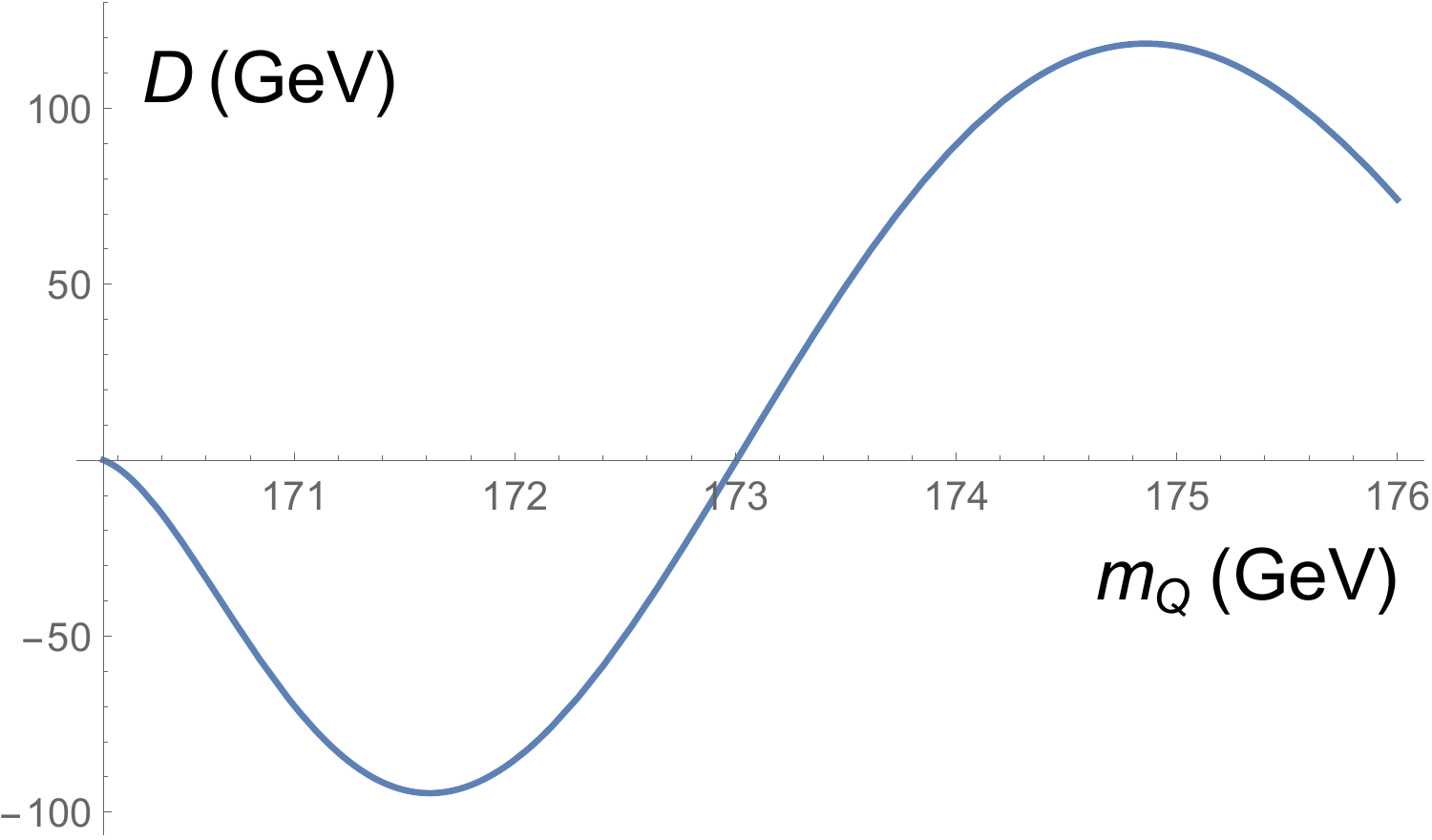}\hspace{1.0 cm} 
\includegraphics[scale=0.2]{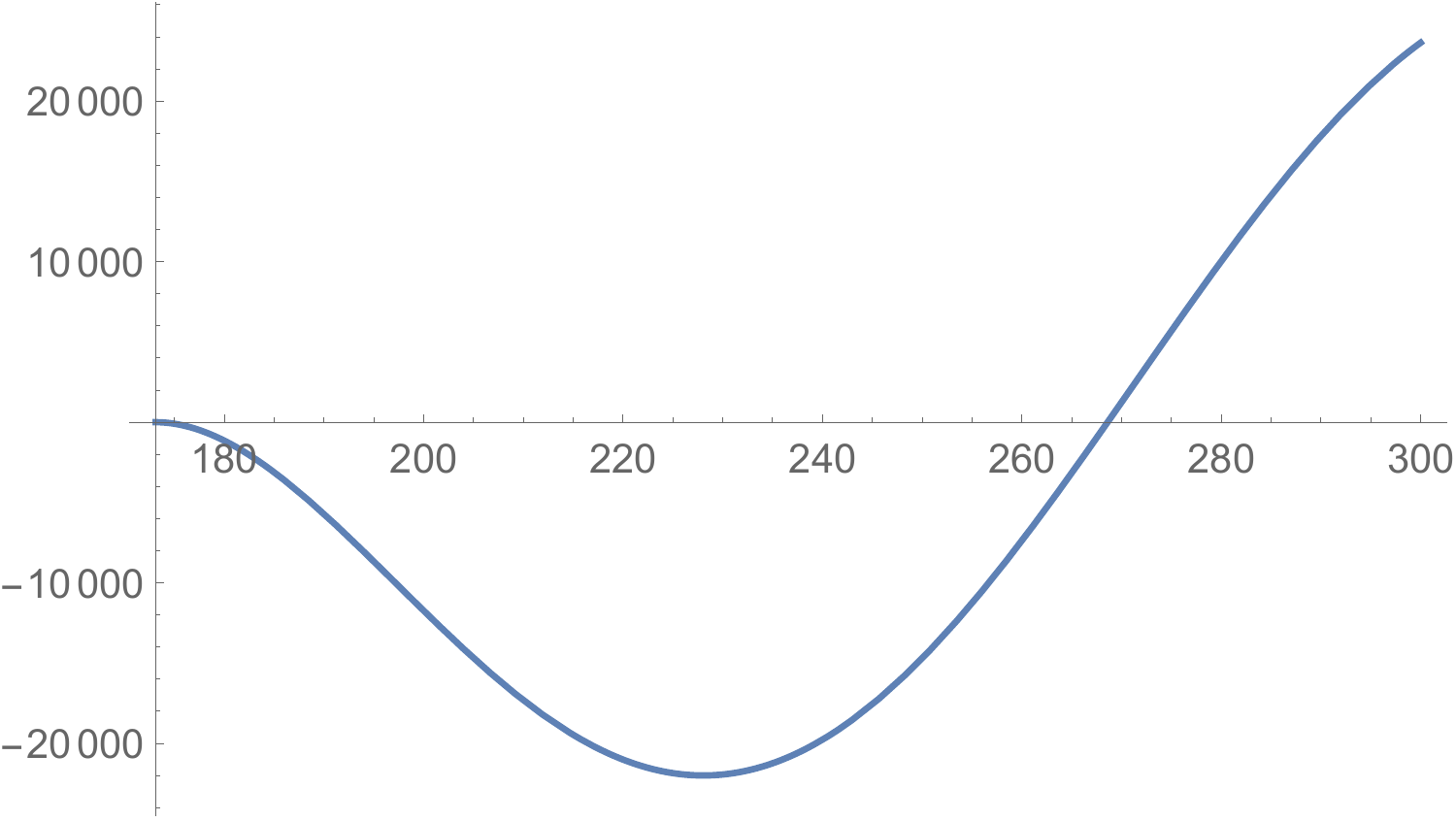}\hspace{1.0 cm}
\includegraphics[scale=0.2]{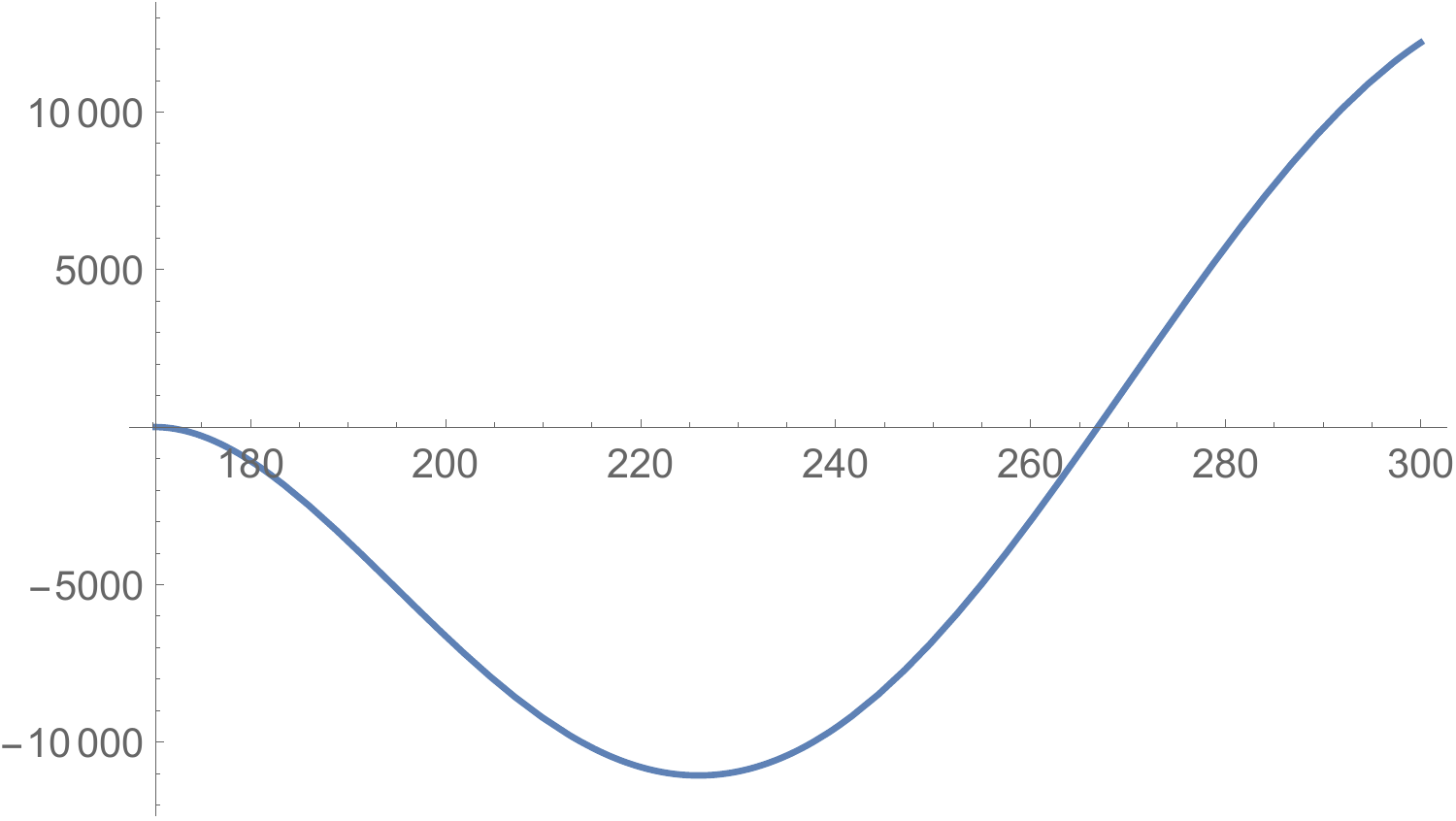}

(a) \hspace{5.5 cm} (b)\hspace{5.5 cm} (c)
\caption{\label{fig2} 
Dependencies of $D(m_Q)$ in Eq.~(\ref{dij}) on $m_Q$ for (a) the $t\to d e^+\nu_4$ decay
with $m_4=170$ GeV, (b) the $L^-\to \nu_1 \bar t d$ decay with $m_t=173$ GeV, and
(c) the $L^-\to \nu_4 \bar u d$ decay with $m_4=170$ GeV.}
\end{center}
\end{figure}

%The unknown function $\Delta\Gamma(m_Q)$, simply converted from $\Delta\rho(m_Q)$, is 
%displayed in Fig.~\ref{fig3}(a) with the obtained $\alpha=2$, $a$ in Eq.~(\ref{dc4}) and 
%$\bar\omega=0.1001$ GeV$^{-1}$, which reveals an oscillatory behavior in $m_Q$ with the 
%first peak being located around $m_Q\approx 173$ GeV, in agreement with the top quark mass 
%$172.69\pm 0.30$ GeV in \cite{PDG}. 

Complying with Eq.~(\ref{dd1}), we work on the derivative 
\begin{eqnarray}
D(m_Q)\equiv \frac{d}{d\omega}
\frac{J_{\alpha}\left(2\omega \sqrt{m_Q^2-m_f^2}\right)}
{J_{\alpha}\left(2\omega\sqrt{m_F^2-m_f^2}\right)}\Bigg|_{\omega=\bar\omega},
\label{dij}
\end{eqnarray}
where the factors independent of $\omega$ have been dropped. The dependence of $D(m_Q)$
on $m_Q\ge m_F$ plotted in Fig.~\ref{fig2}(a) indicates the existence of multiple roots 
for $D(m_Q)=0$. It has been checked that the second derivatives are larger at higher roots 
\cite{Li:2023dqi}, so smaller roots are preferred in order to maximize the stability window 
in $\omega$. Figure~\ref{fig2}(a) reveals that the derivative first vanishes at 
$m_Q\approx 173$ GeV above the boundary $m_Q=m_F$. Namely, $m_4\approx 170$ GeV, leading to 
$m_t\approx 173$ GeV in the dispersive analysis of the $t\to d e^+\nu_4$ decay, can be 
identified as the mass the fourth generation neutrino. 

The uncertainties from the variations of the top quark mass \cite{PDG} and of the scale 
$\mu$ within $[m_Q/2,2m_Q]$ for the coupling constant in Eq.~(\ref{rs}) are negligible; 
they cause less than 1\% error, comparable to the amount from the neglect of the 
$O(\alpha_s)$ constant pieces in Eq.~(\ref{ft1}). The small error means that our result 
is robust against the scale changes and higher-order effects, owing to the large threshold 
mass $m_f$. The dispersive approach itself also bears theoretical uncertainties, 
which mainly arise from the large $N$ assumption. Under this assumption the solution in 
Eq.~(\ref{so3}) is approximated by a single Bessel function, so that the determination 
of the parameter $\bar\omega$ from the fit to the initial condition involves 
some uncertainty. We estimate the impact of this uncertainty by adopting an alternative way 
to fix $\bar\omega$, e.g., by fitting the solution to the initial condition at the midpoint
$m_Q^2=(m_f^2+m_F^2)/2$ of the interval $(m_f^2,m_F^2)$. A different value
$\bar\omega=0.1079$ is obtained, which changes the prediction for $m_4$ by less than 1\%.
In view of the top quark decay width $\Gamma_t=1.42^{+0.19}_{-0.15}$ GeV  \cite{PDG}, the 
choice of the active quark number $n_f=6$ and the corresponding QCD scale parameter 
$\Lambda_{\rm QCD}=0.109$ GeV \cite{Li:2023fim,Deur:2016tte} is acceptable. We have tested 
this set of parameters, and find no effect on the result of $m_4$.

\subsection{$m_L$ from the $L^-\to \nu_1 \bar t d$ Decay}

The mass $m_4\approx 170$ GeV of the fourth generation neutrino suggests that the fourth 
generation charged lepton $L$ may be heavier than a top quark, allowing the determination 
of its mass $m_L$ from the dispersion relation for the $L^-\to \nu_1W^-\to \nu_1 \bar t d$ 
decay with the input $m_t\approx 173$ GeV. The construction of the $L^-\to \nu_1 \bar t d$ 
decay width is similar to that of the $t\to d e^+\nu_4$ decay in the previous subsection; 
we compute the tree-level decay width \cite{Chang:1998ze}, and then attach the QCD 
corrections to the $W^-\to \bar t d$ decay, which are available in the literature 
\cite{Chang:1981qq,Alvarez:1987gi,Gorishnii:1990vf,Surguladze:1990tg,Chetyrkin:1996hm,
Kara:2013dua}. The energy fraction $x_d\equiv 2E_d/m_L$ of the down quark and the 
dimensionless variable $y_W\equiv q^2/m_L^2$ from the invariant mass squared of the $\bar td$ 
pair, i.e., of the $W$ boson, have the ranges the same as in Eq.~(\ref{ran}), but with the 
notation $\eta=m_t^2/m_L^2$. The hard kernel is proportional to
\begin{eqnarray}
(p_L\cdot p_t)(p_d\cdot p_{\nu_1})\propto 
(1+y_W-x_d)(x_d-y_W+\eta),
\end{eqnarray}
where $p_L$ ($p_t$, $p_{\nu_1}$) labels the momentum of the heavy lepton $L$ (the 
anti-top quark $\bar t$, the light neutrino $\nu_1$). 

The resultant $L^-\to \nu_1 \bar t d$ decay width is written as
\begin{eqnarray}
\Gamma^{\rm p}(m_L)&\propto& G_F^2m_L^5(1-\eta)^3\int_0^1dxx
\int_0^1dy\frac{m_W^4[1+y_W(x,y)-(1-\eta) x][(1-\eta)x-y_W(x,y)+\eta]}
{[m_W^2-m_L^2y_W(x,y)]^2+\Gamma_W^4}
\nonumber\\
& &\times\left[1+\frac{\alpha_s(m_L)}{\pi}+\frac{\alpha_s^2(m_L)}
{\pi^2}F_W^{(2)}(x,y)+\frac{\alpha_s^3(m_L)}{\pi^3}F_W^{(3)}(x,y)\right],\label{Lp}
\end{eqnarray}
with $x$ being introduced by the variable change $x_d=(1-\eta)x$ and the expression of 
$y_W$ identical to Eq.~(\ref{yw}). The argument of the running coupling constant has been 
set to $\mu=m_L$. The $O(\alpha_s^2)$ and $O(\alpha_s^3)$ corrections read 
\cite{Chetyrkin:1996hm}
\begin{eqnarray}
F_W^{(2)}(x,y)&=&\frac{365}{24} - 11\zeta(3) + n_f\left[\frac{2}{3}\zeta(3) - \frac{11}{12}\right]
+ \left(\frac{n_f}{6} - \frac{11}{4}\right)\ln[y_W(x,y)],\nonumber\\
F_W^{(3)}(x,y)&=&n_f^2\left[\frac{2}{9}\zeta(3) - \frac{11}{36}\right]\ln[y_W(x, y)] 
+ \frac{n_f^2}{36}\ln^2[y_W(x, y)],\label{fl}
\end{eqnarray}
where $\zeta(3)$ denotes the value of the Riemann zeta function $\zeta(z)$ at $z=3$, and
only the dominant logarithmic terms with the enhancement of $n_f^2$ are kept in the latter 
for simplicity. We point out that the $O(\alpha_s^3)$ piece is not important here, since 
the invariant mass of the $\bar td$ pair must be above the top quark mass. 
Nevertheless, we include it for the $L^-\to \nu_4 \bar u d$ decay to be investigated in the next 
subsection, where the $\bar u d$ quark pair can be generated with a low invariant mass.

We regard the heavy lepton mass $m_L$ as a variable $m_Q$, set the quark-level threshold 
$m_f=m_t$, and choose the integrands in Eq.~(\ref{ij1}) as
\begin{eqnarray}
{\rm Im}\Pi^{({\rm p})}(m^2)&=&\frac{m^5\Gamma^{({\rm p})}(m)}{(m^2-m_f^2)^3},\label{mi2}
\end{eqnarray}
according to the behavior of Eq.~(\ref{Lp}) in the region with $m_Q\sim m_f$,
\begin{eqnarray}
\Gamma^{\rm p}(m_Q)&\propto &\frac{(m_Q^2-m_f^2)^5}{m_Q^5}.\label{asy2}
\end{eqnarray}
Equations~(\ref{mi2}) and (\ref{asy2}) together fix the initial condition of the 
subtracted unknown function in the limit $m_Q\to m_f$,
\begin{eqnarray}
\Delta\rho(m_Q^2)&\propto &(m_Q^2-m_f^2)^2,\label{pt2}
\end{eqnarray}
and the index $\alpha=2$ in Eq.~(\ref{so3}). With the QCD scale parameter 
$\Lambda_{\rm QCD}=0.109$ GeV for the active quark number $n_f=6$ in Eq.~(\ref{rs}), we  
attain the parameter $\bar\omega=0.0155$ GeV$^{-1}$ from the fit of 
$\Delta\rho(m_Q^2)|_{\omega=\bar\omega}$ to the initial condition $-{\rm Im}\Pi^{\rm p}(m_Q^2)$ 
in the interval $(m_f^2,m_F^2)$, $m_F=m_t+m_{\pi^0}$ being the hadronic threshold. 

The dependence of the derivative $D(m_Q)$ in Eq.~(\ref{dij}) on $m_Q$ for $\alpha=2$ and 
$\bar\omega=0.0155$ GeV$^{-1}$ is presented in Fig.~\ref{fig2}(b), where the first root, 
to be identified as the heavy lepton mass $m_L$, appears at $m_Q\approx 269$ GeV. Varying 
the top quark mass \cite{PDG} and the scale $\mu$ for $\alpha_s(\mu)$ within the range 
$[m_Q/2,2m_Q]$ produces only 1\% uncertainty. The uncertainty from the theoretical framework,
as estimated in the previous subsection, is also found to be about 1\%. The mass of a 
sequential heavy charged lepton 
roughly below 100 GeV has been excluded at 95\% CL via measurements of the $L\to\nu W$ decays 
\cite{PDG}. It was proposed to detect heavy charged lepton pairs in a Drell-Yan process at 
LHC \cite{Allanach:2001sd}, whose production rate at $m_L\approx 270$ GeV is about $10^{-3}$ 
lower than the top pair one due to the suppression from the coupling constant and parton 
distribution functions.

\subsection{Cross-check by the $L^-\to \nu_4 \bar u d$ Decay}

We corroborate that the masses of the fourth generation leptons predicted in the previous 
subsections satisfy the dispersion relation for the $L^-\to \nu_4W^-\to \nu_4 \bar u d$ 
decay. The associated formula can also be derived by combining the tree-level 
$L^-\to \nu_4 \bar u d$ decay width and the QCD corrections to the $W^-\to \bar u d$ decay 
width up to three loops \cite{Chetyrkin:1996hm}. The expression is simpler, as the energy 
fraction is assigned to the $\bar u$ quark with $x_u\equiv 2E_u/m_L$, whose range is the same 
as in Eq.~(\ref{ran}). The variable $y_W\equiv q^2/m_L^2$, representing the invariant mass 
squared of the $\bar ud$ pair, takes values within 
\begin{eqnarray}
0\le &y_W& \le \frac{x_u(1-\eta-x_u)}{1-x_u},\label{ran3}
\end{eqnarray}
for the notation $\eta=m_4^2/m_L^2$. The hard kernel is analogous to Eq.~(\ref{h1}), 
\begin{eqnarray}
(p_L\cdot p_u)(p_d\cdot p_{\nu_4})\propto x_u(1-\eta-x_u),
\end{eqnarray}
with the momentum $p_u$ of the $\bar u$ quark. 

The $L^-\to \nu_4 \bar u d$ decay width is given, in a perturbation theory, by
\begin{eqnarray}
\Gamma^{\rm p}&\propto&G_F^2m_L^5(1-\eta)^5\int_0^1dx\frac{x^2(1-x)^2}
{1-(1-\eta)x}\int_0^1\frac{m_W^4dy}{[m_W^2-m_L^2y_W(x,y)]^2+\Gamma_W^4}\nonumber\\
& &\times\left[1+\frac{\alpha_s(m_L)}{\pi}+\frac{\alpha_s^2(m_L)}
{\pi^2}F_W^{(2)}(x,y)+\frac{\alpha_s^3(m_L)}
{\pi^3}F_W^{(3)}(x,y)\right],\label{Lp4}
\end{eqnarray}
where the variable $x$ is related to $x_u$ via $x_u=(1-\eta)x$, and the function
\begin{eqnarray}
y_W(x,y)=\frac{(1-\eta)^2x(1-x)}{1-(1-\eta)x}y,\label{yw4}
\end{eqnarray}
serves as a variable change from $y_W$ to $y$. The $O(\alpha_s^2)$ and $O(\alpha_s^3)$ 
corrections, $F_W^{(2)}(x,y)$ and $F_W^{(3)}(x,y)$, have been defined 
in Eq.~(\ref{fl}). 

Equation~(\ref{Lp4}) behaves like Eq.~(\ref{asy2}) in the threshold region with 
$m_Q\sim m_f=m_4$, so the integrands in Eq.~(\ref{ij1}) follow Eq.~(\ref{mi2}). 
We then have the initial condition of the subtracted unknown function in Eq.~(\ref{pt2}) 
in the limit $m_Q\to m_f$, and the index $\alpha=2$ for Eq.~(\ref{so3}). The parameter 
$\bar\omega=0.0155$ GeV$^{-1}$ is drawn from the fit of 
$\Delta\rho(m_Q^2)|_{\omega=\bar\omega}$ to the initial condition in the interval 
$(m_f^2,m_F^2)$, where the upper bound is chosen as $m_F=m_4+m_{\pi^0}+m_{\pi^-}$, with $n_f=6$
and $\Lambda_{\rm QCD}=0.109$ GeV in Eq.~(\ref{rs}) and the inputs $m_4= 170$ GeV and 
$m_{\pi^-}=0.140$ GeV \cite{PDG}. The dependence of the derivative $D(m_Q)$ in Eq.~(\ref{dij}) 
on $m_Q$ for $\alpha=2$ and $\bar\omega=0.0155$ GeV$^{-1}$ is shown in Fig.~\ref{fig2}(c), 
where the first root located at $m_Q\approx 267$ GeV is identified as the heavy lepton mass 
$m_L$. This value is very close to $m_L\approx 269$ GeV in the previous subsection with deviation 
less than 1\%. The concordance between the analyses of different modes supports the 
consistency of our formalism and the predictions for the fourth generation lepton masses. The 
variation of the scale $\mu$ for $\alpha_s(\mu)$ is the major source of the theoretical
uncertainty, resulting in $m_L\approx 248$ GeV for $\mu=m_Q/2$ and $m_L\approx 289$ GeV for 
$\mu=2m_Q$, i.e., $m_L\approx (267\pm 20)$ GeV. The sensitivity originates from the lower 
invariant mass that the $\bar u d$ quark pair can reach as mentioned before, and from the
resultant enhancement of the radiative logarithmic corrections. The uncertainty from the 
theoretical framework may amount to 6\% in this decay channel. It is noticed that 
the neglect of the $O(\alpha_s^3)$ piece $F_W^{(3)}$ would increase the result of $m_L$ 
by 20\%. It highlights the relevance of this higher-order contribution here, contrary to 
the $L^-\to \nu_1 \bar t d$ case.

%arXiv:2312.09099 (replaced) [pdf, ps, html, other]
%The oblique parameters from arbitrary new fermions
%Francisco Albergaria, Darius Jurčiukonis, Luís Lavoura
 
\section{EXPERIMENTAL CONSTRAINTS ON THE FOURTH-GENERATION LEPTONS}

The experimental bounds on the parameters in the sequential fourth generation model
have been surveyed in the literature. Our predictions for $m_4\approx 170$ GeV and 
$m_L\approx 270$ GeV, and for their mass splitting accommodate the fit results 
in \cite{Bellantoni:2012ag}, but are higher than $m_4=58.0$ GeV and $m_L=113.6$ 
GeV from \cite{Eberhardt:2012sb}, which assumed lighter fourth generation quarks of masses 
about 600 GeV. Unlike the fourth generation quarks with masses above a TeV scale 
\cite{Li:2023fim}, the fourth generation leptons do not form bound states in a Yukawa 
potential. Though they are not involved in the Higgs production via gluon fusion, the 
charged lepton contributes to the Higgs decay into a photon pair, $H\to \gamma\gamma$. 
Therefore, the argument for the fourth generation quarks to bypass the experimental 
constraints from the Higgs decay by means of the formation of bound states \cite{Li:2023fim} 
does not apply. It is necessary to scrutinize whether the Higgs decay data rule out the 
existence of the fourth generation charged lepton. 

The $H\to \gamma\gamma$ decay width with the contribution from the fourth generation charged 
lepton is expressed as \cite{Ishiwata:2011hr}
\begin{eqnarray}
\Gamma_{\gamma\gamma}=\frac{\alpha_e^2 G_F m_H^3}{128\sqrt{2}\pi^3}|J_{\gamma\gamma}|^2,
\end{eqnarray}
where $\alpha_e$ is the fine structure constant and $m_H$ the Higgs mass. The decay amplitude 
contains three terms,
\begin{eqnarray}
J_{\gamma\gamma}=\left(\frac{2}{3}\right)^2N_cI(r_t)+I(r_L)+K(r_W),\label{gaga}
\end{eqnarray}
with the functions
\begin{eqnarray}
I(x)&=&\frac{2}{x^2}[x + (x - 1)f(x)],\;\;\;\;
K(x)= -\left[2+\frac{3}{x} + \frac{3}{x^2}(2x - 1)f(x)\right],\nonumber\\
f(x)&=&\arcsin^2\sqrt{x}.
\end{eqnarray}
Inserting $r_{t,L,W}=m_H^2/(4m_{t,L,W}^2)$ with $m_H= 125$ GeV, 
$m_t= 173$ GeV, $m_L= 270$ GeV and $m_W= 80.4$ GeV, we obtain the ratio
\begin{eqnarray}
\frac{|J_{\gamma\gamma}|^2}{|J_{\gamma\gamma}^{\rm SM}|^2}= 0.63,
\end{eqnarray}
where $J_{\gamma\gamma}^{\rm SM}$ comes from Eq.~(\ref{gaga}) but without the $I(r_L)$ piece.
That is, $\Gamma_{\gamma\gamma}$ is reduced by 37\%; the terms from fermion loops in 
Eq.~(\ref{gaga}) are destructive to the dominant one from a $W$-boson loop, 
so the fourth generation charged lepton decreases the decay width. 

The recent calculation for the above width, which takes into account various sources of 
subleading corrections in the SM, gives 
$\Gamma_{\gamma\gamma}^{\rm SM}=(9.28\pm 0.16)\times 10^{-6}$ GeV \cite{Davies:2021zbx}.
The theoretical uncertainty covers those from QCD contributions, electroweak contributions 
and variations of input parameters. We mention another result 
$\Gamma_{\gamma\gamma}^{\rm SM}\approx 9.4\times 10^{-6}$ GeV with similar precision
\cite{Sturm:2014nva}. As to the data, we estimate 
$\Gamma_{\gamma\gamma}^{\rm data}\approx (9.3^{+4.8}_{-3.5})\times 10^{-3}$ MeV from the 
width $3.7^{+1.9}_{-1.4}$ MeV of a Higgs boson and the measured $H\to\gamma\gamma$ branching 
ratio $(2.50\pm 0.20)\times 10^{-3}$ \cite{PDG}, where the small error from the branching 
ratio has been ignored. Hence, the 37\% deduction of the SM prediction \cite{Davies:2021zbx}, 
i.e., the modified width $\Gamma_{\gamma\gamma}= (5.85\pm 0.10)\times 10^{-3}$ MeV 
still agrees with the data within the experimental uncertainty. We postulate that the Higgs 
decay data does not exclude the existence of the fourth generation charged lepton. 

Together with the masses of the fourth generation quarks predicted in \cite{Li:2023fim}, we 
are ready to address the fourth generation contributions to the oblique parameters $S$, $T$ 
and $U$. It has been known that the superheavy fourth generation quarks form bound states 
under a strong Yukawa potential \cite{Hung:2009hy,Enkhbat:2011vp}. The oblique parameters 
are ultraviolet finite and do not probe the internal structure of these bound 
states, similar to the cases of the Higgs production from gluon fusion and the Higgs decay 
into a photon pair \cite{Li:2023fim}. Therefore, the contribution to the oblique parameters 
from the fourth generation quarks can be assessed in the same effective approach 
as in \cite{Li:2023fim}. The fourth generation leptons are too light to form bound states, 
so we quantify their impact by adopting the one-loop formulas in \cite{He:2001tp}. 
It will be explained in the next section that the mixing of the fourth generation leptons with 
the other light generations is quite weak, an observation in line with the assumption made in 
\cite{He:2001tp}. Substituting $m_4=170$ GeV, $m_L=270$ GeV and the $Z$ boson mass $m_Z=91.2$ 
GeV \cite{PDG} into Eqs.~(9)-(11) in \cite{He:2001tp}, we get straightforwardly
\begin{eqnarray}
S=0.003,\;\;\;\;T=0.178,\;\;\;\;U=0.009.\label{opp}
\end{eqnarray}
It is noticed that the fourth generation leptons with the distinct masses $m_4$ and $m_L$ 
induce the oblique parameter $T$ more significantly, which measures isospin 
breaking effects of new physics \cite{Peskin:1991sw}.

Below we estimate the order of magnitude of the contributions from the bound states formed
by the fourth generation quarks, and certify that they are much smaller than from the 
fourth generation leptons. In the viewpoint of an effective theory, the oblique parameters 
describe the process, where an electroweak boson $W$ turns into a vector boson $V$, 
a bound state of the fourth generation quarks, through a dimensionless coupling $g_{WV}$ 
between them. The vector boson $V$ propagates according to a Breit-Wigner factor 
$1/(m_V^2-s-i\sqrt{s}\Gamma_V)$ with the invariant mass squared $s=q^2$, where $m_V$ 
($\Gamma_V$, $q$) is the mass (width, momentum) of $V$. The vector boson $V$ then 
transforms into another electroweak boson $W'$ with the dimensionless coupling $g_{W'V}$. 
The corresponding amplitude is thus formulated, in the effective approach, as
\begin{eqnarray}
\frac{g_{WV}g_{W'V}s(g^{\mu\nu}-q^\mu q^\nu/s)}{m_V^2-s-i\sqrt{s}\Gamma_V},\label{amp}
\end{eqnarray}
where factors irrelevant to our reasoning are implicit. 

Take the contribution from a $\bar b' b'$ bound state to the oblique parameter $S$ as an 
example. The product of the effective couplings $g_{WV}g_{W'V}$ can be fixed by matching 
the amplitude in Eq.~(\ref{amp}) to the perturbative one $S_{b'}(s)$ in the fundamental 
theory at a high scale $s$ \cite{He:2001tp},
\begin{eqnarray}
S_{b'}(s) &=& \frac{N_c}{6\pi}\left\{ 2(3-4Y_{b'})\frac{m_{b'}^2}{s} 
+2Y_{b'}\ln\frac{m_{b'}^2}{s} +
\left[\left(\frac{3}{2} - 2Y_{b'}\right)\frac{m_{b'}^2}{s} 
- Y_{b'}\right]G\left(\frac{m_{b'}^2}{s}\right)\right\},\label{sbp}\\
G(x)&=&-4\sqrt{4x - 1}\arctan\frac{1}{\sqrt{4x - 1}},\nonumber
\end{eqnarray}
$N_c=3$ being the color number and $Y_{b'}=1/6$ the hypercharge of a $b'$ quark.
Relating Eq.~(\ref{amp}) to Eq.~(\ref{sbp}) at the large mass squared $s\approx m_{b'}^2$, 
we have
\begin{eqnarray}
\frac{g_{WV}g_{W'V}m_{b'}^2}{m_V^2-m_{b'}^2}\sim S_{b'}(m_{b'}^2),\label{eff}
\end{eqnarray}
where the product $m_{b'}\Gamma_V< m_V^2-m_{b'}^2 $ has been neglected in the denominator
for $m_V\sim 3$ TeV and $\Gamma_V\sim 500$ GeV from \cite{Li:2023fim}.
The above relation then implies $g_{WV}g_{W'V}\sim S_{b'}(m_{b'}^2)(m_V^2/m_{b'}^2-1)$.

%\begin{eqnarray}
%\frac{g_{WV}g_{W'V}m_{b'}^2}{\Gamma_V}\sim \left|S_{b'}\right|,%\label{eff}
%\end{eqnarray}
%$g_{WV}g_{W'V}\sim S_{b'}\Gamma_V/m_V$.
%at the scale $s\approx m_V^2$. 
%The absolute values are considered, because signs are not relevant for demonstrating
%the smallness of this contribution
%We have evaluated the masses and widths of the $\bar b'b'$ bound states in both 
%nonrelativistic and relativistic approaches, and found that they take the values above 
%3 TeV and about 500 GeV \cite{Li:2023fim}, respectively. 

Extrapolating Eq.~(\ref{amp}) to the region with $s\approx m_Z^2\ll m_V^2$, we obtain 
the suppression factor on the vector boson contribution relative to the perturbative one
$S_{b'}(m_Z^2)$,
\begin{eqnarray}
\frac{g_{WV}g_{W'V}s}{S_{b'}(m_Z^2)(m_V^2-s)}\approx 
\frac{S_{b'}(m_{b'}^2) m_Z^2(m_V^2-m_{b'}^2)}{S_{b'}(m_Z^2)m_{b'}^2 m_V^2} 
\sim 10^{-4}.\label{per}
\end{eqnarray}
The exact values of $S_{b'}(m_Z^2)$ and $S_{b'}(m_{b'}^2)$ do not matter actually, for
they are of the same order of magnitude, and largely cancel in the above ratio. 
Equation~(\ref{per}) manifests that the vector contribution decreases like $m_Z^2/m_V^2$
($m_{b'}$ and $m_V$ are of the same order of magnitude), and that the $\bar b't'$ and 
$\bar t't'$ contributions are even more suppressed by huge vector boson masses. With 
$S_{b'}(m_Z^2)\approx 0.5$ from Eq.~(\ref{sbp}), it is easy to see that the $\bar b' b'$ 
contribution is of $O(10^{-5})$, lower than the one from the fourth generation 
lepton by a factor of $10^{-2}$. The same observation applies to the bound-state 
contributions to the oblique parameters $T$ and $U$, and will not be repeated here.

We conclude that the contributions from the fourth generation fermions to the oblique
parameters are dominated by those from the leptons already given in Eq.~(\ref{opp}). The 
global fits to electroweak precision measurements indicate that new physics effects 
cause the oblique parameters \cite{PDG}
\begin{eqnarray}
S =-0.04\pm 0.10,\;\;\;\;T =0.01\pm 0.12,\;\;\;\;U =-0.01\pm 0.09.
\end{eqnarray}
Equation~(\ref{opp}), consistent with the above data within $2\sigma$ error, 
i.e., 95\% CL, asserts that the sequential fourth generation model is not ruled out, and 
deserves further pursuits.

\section{CONSTRAINTS ON THE PMNS MATRIX ELEMENTS}

We have explored the dispersion relations for the mixing of neutral leptonic states
$\mathcal{L}^-\ell^+$ and $\mathcal{L}^+\ell^-$, like $\mu^-e^+$ and $\mu^+ e^-$, in 
\cite{Li:2023ncg,Li:2024awx}, where $\mathcal{L}$ ($\ell$) stands for a massive (light) 
charged lepton, and observed that these relations impose strong constrains 
on the neutrino masses and the PMNS matrix elements. If the 
fourth generation leptons exist, they might affect the constraints inferred from the 
three-generation case. The only assumption underlying the analysis is that the electroweak 
symmetry of the SM is restored at a high energy scale $\Lambda$ 
\cite{Chien:2018ohd,Huang:2020iya}, which could be realized in, for instance, the 
composite Higgs model \cite{Kaplan:1983fs}. It was then argued based on the unitarity of 
the PMNS matrix that the mixing phenomenon disappears, as the electroweak symmetry of 
the SM is restored. The disappearance of the mixing at $m_\mathcal{L}>\Lambda$ was 
taken as the input to the dispersion relation, $m_\mathcal{L}$ being the mass of the 
heavy lepton $\mathcal{L}$, and the solution at low $m_\mathcal{L}<\Lambda$, 
i.e., in the symmetry broken phase, was found to bind the neutrino masses and 
the PMNS matrix elements involved in the mixing amplitude \cite{Li:2023ncg}. We point out 
that $\Lambda$ plays a role similar to the large scale in Sec.~III, above 
which perturbative inputs are reliable.

We decompose the mixing amplitude $\Pi(m_Q^2)$ into a sum over various intermediate 
neutrino channels,
\begin{eqnarray}
\Pi(m_\mathcal{L}^2)&=& M(m_\mathcal{L}^2)-\frac{i}{2}\Gamma(m_\mathcal{L}^2)\nonumber\\
&\equiv& \sum_{i,j=1}^4\lambda_{i}\lambda_{j}\left[M_{ij}(m_\mathcal{L}^2)
-\frac{i}{2}\Gamma_{ij}(m_\mathcal{L}^2)\right],
\end{eqnarray} 
where $\lambda_i\equiv U^*_{\mathcal{L} i}U_{\ell i}$ is the product of the $4\times 4$ PMNS 
matrix elements. It has been elaborated \cite{Li:2024awx} that the dependence of the real part 
$M_{ij}(m_\mathcal{L}^2)$ on the intermediate neutrino masses, i.e., the contribution which 
survives the summation over all intermediate channels, begins at the three-loop level in 
the symmetric phase. In other words, the argument on the disappearance of the mixing 
phenomenon, $\sum_{i,j}\lambda_{i}\lambda_{j}M_{ij}(m_\mathcal{L}^2)\approx 0$, in the 
symmetric phase is valid to high accuracy. The corresponding dispersion relation is thus 
written as
\begin{eqnarray}
M(m_\mathcal{L}^2)=\frac{1}{2\pi}\int^{\Lambda^2} dm^2
\frac{\Gamma(m^2)}{m_\mathcal{L}^2-m^2}\approx 0,\label{dis}
\end{eqnarray}
for $m_\mathcal{L}>\Lambda$.

To diminish the dispersive integral in Eq.~(\ref{dis}) for arbitrary $m_\mathcal{L}>\Lambda$, 
some conditions must be met by the PMNS matrix elements. We quote the asymptotic behavior 
of the imaginary part $\Gamma_{ij}(m^2)$ for $m<\Lambda$ \cite{Li:2023ncg,Cheng,BSS}
\begin{eqnarray}
\Gamma_{ij}(m^2)\approx \Gamma^{(1)}_{ij}m^2+\Gamma^{(0)}_{ij}
+\frac{\Gamma^{(-1)}_{ij}}{m^2}+\cdots,\label{exp}
\end{eqnarray}
with the coefficients
\begin{eqnarray}
\Gamma^{(1)}_{ij}&=&\frac{4m_W^4-6m_W^2(m_i^2+m_j^2)+m_i^2m_j^2}
{2(m_W^2-m_i^2)(m_W^2-m_j^2)},\nonumber\\
\Gamma^{(0)}_{ij}&=&-\frac{3(m_i^2+m_j^2)\left[4m_W^4-4m_W^2(m_i^2+m_j^2)+m_i^2m_j^2\right]}
{2(m_W^2-m_i^2)(m_W^2-m_j^2)},\nonumber\\
\Gamma^{(-1)}_{ij}&=&\frac{3(m_i^4+m_j^4)\left[4m_W^4-2m_W^2(m_i^2+m_j^2)+m_i^2m_j^2\right]}
{2(m_W^2-m_i^2)(m_W^2-m_j^2)}.\label{mij}
\end{eqnarray}
The terms $\Gamma_{ij}^{(n)}$ give contributions to the dispersive integral in 
Eq.~(\ref{dis}), which scale like $\Lambda^4$, $\Lambda^2$ and $\ln\Lambda^2$ for $n=1,0$ 
and $-1$, respectively. Since $\Lambda$ just needs to be of order of the symmetry 
restoration scale, instead of a definite value, it is unlikely that the above huge 
contributions happen to cancel among themselves. The finiteness of the dispersive integral 
can only be achieved by requiring
\begin{eqnarray}
\sum_{i,j=1}^4\lambda_{i}\lambda_{j} \Gamma_{ij}^{(n)}\approx 0,\;\;\;\;n=1,0,-1.\label{cons}
\end{eqnarray}

Once the conditions in Eq.~(\ref{cons}) are fulfilled, we recast the dispersive integral 
in Eq.~(\ref{dis}) into
\begin{eqnarray}
\int^{\Lambda^2} dm^2\frac{\Gamma(m^2)}{m_\mathcal{L}^2-m^2}\approx
\frac{1}{m_\mathcal{L}^2}\sum_{i,j=1}^4\lambda_{i}\lambda_{j}g_{ij},\label{cong}
\end{eqnarray}
with the factors
\begin{eqnarray}
g_{ij}\equiv\int_{t_{ij}}^\infty dm^2\left[\Gamma_{ij}(m^2)-
\Gamma_{ij}^{(1)}m^2-\Gamma^{(0)}_{ij}-\frac{\Gamma^{(-1)}_{ij}}{m^2}\right],\label{gi}
\end{eqnarray}
and the thresholds $t_{ij}=(m_i+m_j)^2$. The approximation 
$1/(m_\mathcal{L}^2-m^2)\approx 1/m_\mathcal{L}^2$ has been made for large 
$m_\mathcal{L}>\Lambda$, because the integral receives contributions only 
from finite $m<\Lambda$. The integrand in the square brackets decreases like $1/m^4$, so 
the upper bound of $m^2$ in Eq.~(\ref{gi}) can be pushed to infinity safely. We place 
the final condition, labeled by $n=f$,
\begin{eqnarray}
\sum_{i,j=1}^4\lambda_{i}\lambda_{j}g_{ij}\approx 0,\label{gij}
\end{eqnarray}
to ensure the almost nil dispersive integral. That is, the realization of Eqs.~(\ref{cons}) 
and (\ref{gij}) establishes a solution to the integral equation in Eq.~(\ref{dis}). 

We employ the unitarity condition to eliminate $\lambda_4=-\lambda_1-\lambda_2-\lambda_3$, 
and define the ratios of the PMNS 
matrix elements,
\begin{eqnarray}
r_1\equiv\frac{U^*_{\mathcal{L} 1}U_{e1}}{U^*_{\mathcal{L} 2}U_{e2}}\equiv u_1+iv_1,\;\;\;\;
r_3\equiv\frac{U^*_{\mathcal{L} 3}U_{e3}}{U^*_{\mathcal{L} 2}U_{e2}}\equiv u_3+iv_3,
\end{eqnarray}
where $\mathcal{L}$ represents either $\mu$ or $\tau$, and the unknowns $u_{1,3}$ and 
$v_{1,3}$ will be constrained below. The $n=1,0$ and $-1$ conditions can be factorized into
\begin{eqnarray}
& &\left(r_1\frac{m_4^2-m_1^2}{m_W^2-m_1^2}+\frac{m_4^2-m_2^2}{m_W^2-m_2^2}
+r_3\frac{m_4^2-m_3^2}{m_W^2-m_3^2}\right)^2
\approx 0,\label{cs1}\\
& &\left(r_1\frac{m_4^2-m_1^2}{m_W^2-m_1^2}+\frac{m_4^2-m_2^2}{m_W^2-m_2^2}
+r_3\frac{m_4^2-m_3^2}{m_W^2-m_3^2}\right)
\left(r_1\frac{m_4^2-m_1^2}{m_W^2-m_1^2}P_1+\frac{m_4^2-m_2^2}{m_W^2-m_2^2}P_2
+r_3\frac{m_4^2-m_3^2}{m_W^2-m_3^2}P_3\right)
\approx 0,\label{cs2}\\
& &\left(r_1\frac{m_4^2-m_1^2}{m_W^2-m_1^2}+\frac{m_4^2-m_2^2}{m_W^2-m_2^2}
+r_3\frac{m_4^2-m_3^2}{m_W^2-m_3^2}\right)
\left(r_1\frac{m_4^2-m_1^2}{m_W^2-m_1^2}Q_1+\frac{m_4^2-m_2^2}{m_W^2-m_2^2}Q_2
+r_3\frac{m_4^2-m_3^2}{m_W^2-m_3^2}Q_3\right)\approx 0,\label{cs3}
\end{eqnarray}
respectively, with the functions
\begin{eqnarray}
P_i&=&\frac{m_W^2(m_4^2+m_i^2)-m_4^2m_i^2}{m_4^4}\nonumber\\
Q_i&=&\frac{2m_W^4(m_4^2+m_i^2)+m_4^2m_i^2(m_4^2+m_i^2)-m_W^2(m_4^4+3m_4^2m_i^2+m_i^4)}{m_4^6}.
\label{pq}
\end{eqnarray}
The above conditions have been presented in terms of dimensionless mass ratios as 
done in \cite{Li:2023ncg}.

We observe immediately that Eqs.~(\ref{cs1})-(\ref{cs3}) can be respected simultaneously
by diminishing the common factor
\begin{eqnarray}
r_1\frac{m_4^2-m_1^2}{m_W^2-m_1^2}+\frac{m_4^2-m_2^2}{m_W^2-m_2^2}
+r_3\frac{m_4^2-m_3^2}{m_W^2-m_3^2}\approx 0,\;\;
{\rm i.e.},\;\; \sum_{i=1}^3U^*_{\mathcal{L} i}U_{ei}=-U^*_{\mathcal{L} 4}U_{e4}
\sim O\left(\frac{m_i^2}{m_{4,W}^2}\right).
\label{au}
\end{eqnarray}
It means that the unitarity of the $3\times 3$ PMNS matrix holds well, up to 
$O(m_i^2/m_{4,W}^2)$ corrections, when the mass $m_4$ of the fourth generation neutrino 
is of order of the $W$ boson mass $m_W$. In the three-generation case there are only two 
unknowns $u_1$ and $v_1$ \cite{Li:2023ncg}, and Eq.~(\ref{au}) reduces to
\begin{eqnarray}
r_1\frac{m_3^2-m_1^2}{m_W^2-m_1^2}+\frac{m_3^2-m_2^2}{m_W^2-m_2^2}\approx 0,\;\;
{\rm i.e.},\; \sum_{i=1}^2U^*_{\mathcal{L} i}U_{ei}=-U^*_{\mathcal{L} 3}U_{e3}
\sim O\left(\frac{m_i^2}{m_{3}^2}\right).\label{au3}
\end{eqnarray}
Namely, we would have the order-of-magnitude estimate 
$|U^*_{\mu 3}U_{e3}|\sim \sin\theta_{23}\cos\theta_{13}\sin\theta_{13}\sim m_2^2/m_3^2\approx 0.03$
for $\mathcal{L}=\mu$, much lower than the measured value $0.11$ \cite{PDG}. The above 
argument is applicable to the quark mixing, such as the $c\bar u$-$\bar cu$ mixing, 
simply by substituting the quark masses $m_{d,s,b,b'}$ for the neutrino masses $m_{1,2,3,4}$. 
The unitarity of the $3\times 3$ CKM matrix is then expected to be violated at the level 
$m_b^2/m_W^2\approx 10^{-3}$, in agreement with the conclusion from recent global fits to 
precision measurements in flavor physics \cite{Seng:2021gmh,Kitahara:2023xab,Kitahara:2024azt}; 
in the three-generation case, we would have the order-of-magnitude estimate 
$V^*_{cb}V_{ub}\sim m_s^2/m_b^2\approx 6.3\times 10^{-4}$, much higher than the 
measured value $1.5\times 10^{-4}$ \cite{PDG}. In this sense, the data for the PMNS 
and CKM matrix elements have hinted the existence of the sequential fourth generation 
fermions according to the dispersive constraints.

%The right-handed fourth-generation neutrino 
%has an extremely weak mixing with the other generations of leptons, and behave like a 
%sterile neutrino. Moreover, it does not decay into a heavier fourth-generation charged lepton, 
%so it should be very stable. In this sense a fourth generation neutrino could be a candidate 
%of dark matter. This subject will be investigated in detail elsewhere. 

The $n=f$ condition in Eq.~(\ref{gij}) possesses a complicated and lengthy form. To an 
excellent accuracy, we concentrate on the leading terms of a piece
\begin{eqnarray}
G_{ij}=g_{ij}-g_{i4}-g_{j4}+g_{44},\label{fij}
\end{eqnarray}
in the sum over the intermediate channels, which is, after the elimination of $\lambda_4$, 
expanded in powers of $m_i/m_{4,W}$,
\begin{eqnarray}
G_{ij}\approx\frac{-20m_W^4m_4^4+28m_W^2m_4^6-7m_4^8+(16m_W^4m_4^3-28m_W^2m_4^5+ 
12m_4^7)(m_i+m_j)}{2(m_W^2-m_4^2)^2(m_W^2-m_i^2)(m_W^2-m_j^2)}.
\end{eqnarray}
Equation~(\ref{gij}) then turns into
\begin{eqnarray}
\sum_{i,j=1}^3 G_{ij}&\approx& -(20m_W^4-28m_W^2m_4^2+7m_4^4)G_0
+(16m_W^4-28m_4^2m_W^2+ 12m_4^4)G_1 \approx 0,\nonumber\\
G_0&=&\left(\frac{r_1}{m_W^2-m_1^2}+\frac{1}{m_W^2-m_2^2}
+\frac{r_3}{m_W^2-m_3^2}\right)^2,\label{fs0}\\
G_1&=&\left(\frac{r_1}{m_W^2-m_1^2}+\frac{1}{m_W^2-m_2^2}
+\frac{r_3}{m_W^2-m_3^2}\right)\left[\frac{r_1 m_1}{m_4(m_W^2-m_1^2)}
+\frac{m_2}{m_4(m_W^2-m_2^2)}+\frac{r_3 m_3}{m_4(m_W^2-m_3^2)}\right].\label{fs1}
\end{eqnarray}

Reexpressing the factors $P_i$ in Eq.~(\ref{pq}) as
\begin{eqnarray}
P_i\approx \frac{m_W^2}{m_4^2}\left(1+\frac{m_i^2}{m_4^2}-\frac{m_i^2}{m_4^2}\right),
\end{eqnarray}
we realize that the $n=0$ condition represents an $O(m_i^2/m_{4,W}^2)$ correction to the $n=1$ 
one. The $n=-1$ condition also appears as an $O(m_i^2/m_{4,W}^2)$ correction to the $n=1$ 
one, so both Eqs.~(\ref{cs2}) and (\ref{cs3}) can be dropped in the search for the solutions
to the dispersive constraints. A distinction from the three-generation case \cite{Li:2023ncg} 
is that the above power corrections become $O(m_{1,2}^2/m_3^2)$, much greater than 
$O(m_{1,2}^2/m_{4,W}^2)$, and the $n=0$ and $-1$ conditions force nontrivial constraints on 
the PMNS matrix elements. Apparently, Eq.~(\ref{fs0}) also represents an $O(m_i^2/m_4^2)$ 
correction to the $n=1$ condition in Eq.~(\ref{cs1}), whose contribution is expected to be 
negligible. The minimization of the four conditions labeled by $n=1,0,-1$ and $f$ was performed 
numerically by tuning the two unknowns $u_1$ and $v_1$ in the three-generation case 
\cite{Li:2023ncg}. We have four unknowns with four constraints from Eqs.~(\ref{cs1}) and 
(\ref{fs1}) here, which contain real and imaginary parts, so that the solution need to be 
constructed in a different way.

%The real and imaginary parts of Eq.~(\ref{fs0}) read
%\begin{eqnarray}
%& &\left(\frac{u_1}{m_W^2-m_1^2}+\frac{1}{m_W^2-m_2^2}
%+\frac{u_3}{m_W^2-m_3^2}\right)^2-
%\left(\frac{v_1}{m_W^2-m_1^2}+\frac{v_3}{m_W^2-m_3^2}\right)^2
%\approx 0,\label{s0r}\\
%& &\left(\frac{u_1}{m_W^2-m_1^2}+\frac{1}{m_W^2-m_2^2}
%+\frac{u_3}{m_W^2-m_3^2}\right)\left(\frac{v_1}{m_W^2-m_1^2}+\frac{v_3}{m_W^2-m_3^2}\right)
%\approx 0,\label{s0i}
%\end{eqnarray}
%which differ from Eqs.~(\ref{nr}) and (\ref{ni}), respectively, by $O(m_i^2/m_4^2)$. 
%Similarly, the vanishing of Eq.~(\ref{fs1}) yields the real and imaginary parts

The real and imaginary parts of Eq.~(\ref{cs1}) lead to
\begin{eqnarray}
& &u_1\frac{m_4^2-m_1^2}{m_W^2-m_1^2}+\frac{m_4^2-m_2^2}{m_W^2-m_2^2}
+u_3\frac{m_4^2-m_3^2}{m_W^2-m_3^2}\approx 0,\label{nr}\\
& &v_1\frac{m_4^2-m_1^2}{m_W^2-m_1^2}
+v_3\frac{m_4^2-m_3^2}{m_W^2-m_3^2}\approx 0,\label{ni}
\end{eqnarray}
from which we derive the expressions for $u_1$ and $v_1$ in terms of $u_3$ and $v_3$,
respectively. The real and imaginary parts of Eq.~(\ref{fs1}) are written as
\begin{eqnarray}
%& &\left(u_1\frac{m_4^2-m_1^2}{m_W^2-m_1^2}+\frac{m_4^2-m_2^2}{m_W^2-m_2^2}
%+u_3\frac{m_4^2-m_3^2}{m_W^2-m_3^2}\right)^2-
%\left(v_1\frac{m_4^2-m_1^2}{m_W^2-m_1^2}
%+v_3\frac{m_4^2-m_3^2}{m_W^2-m_3^2}\right)^2\approx 0,\label{nr}\\
%& &\left(u_1\frac{m_4^2-m_1^2}{m_W^2-m_1^2}+\frac{m_4^2-m_2^2}{m_W^2-m_2^2}
%+u_3\frac{m_4^2-m_3^2}{m_W^2-m_3^2}\right)
%\left(v_1\frac{m_4^2-m_1^2}{m_W^2-m_1^2}
%+v_3\frac{m_4^2-m_3^2}{m_W^2-m_3^2}\right)\approx 0,\label{ni}\\
G_{1r}&\equiv&\left(\frac{u_1}{m_W^2-m_1^2}+\frac{1}{m_W^2-m_2^2}
+\frac{u_3}{m_W^2-m_3^2}\right)\left[\frac{u_1 m_1}{m_4(m_W^2-m_1^2)}
+\frac{m_2}{m_4(m_W^2-m_2^2)}+\frac{u_3 m_3}{m_4(m_W^2-m_3^2)}\right]\nonumber\\
& &-\left(\frac{v_1}{m_W^2-m_1^2}+\frac{v_3}{m_W^2-m_3^2}\right)
\left[\frac{v_1 m_1}{m_4(m_W^2-m_1^2)}
+\frac{v_3 m_3}{m_4(m_W^2-m_3^2)}\right]\approx 0,\label{s1r}\\
G_{1i}&\equiv&\left(\frac{u_1}{m_W^2-m_1^2}+\frac{1}{m_W^2-m_2^2}
+\frac{u_3}{m_W^2-m_3^2}\right)\left[\frac{v_1 m_1}{m_4(m_W^2-m_1^2)}
+\frac{v_3 m_3}{m_4(m_W^2-m_3^2)}\right]\nonumber\\
& &+\left[\frac{u_1 m_1}{m_4(m_W^2-m_1^2)}
+\frac{m_2}{m_4(m_W^2-m_2^2)}+\frac{u_3 m_3}{m_4(m_W^2-m_3^2)}\right]
\left(\frac{v_1}{m_W^2-m_1^2}
+\frac{v_3}{m_W^2-m_3^2}\right)\approx 0.\label{s1i}
\end{eqnarray}
For a given $v_1$, $G_{1r}$ represents a parabola in $u_1$, so Eq.~(\ref{s1r}) 
specifies two roots of $u_1$, which will be assigned as the solutions for the $\mu e$ and 
$\tau e$ mixings separately. On the other hand, $G_{1i}$ describes a straight 
line, and Eq.~(\ref{s1i}) gives a single root of $u_1$. This root is independent 
of $v_1$, such that no definite value of $v_1$ can be associated with it to form a solution.

We assume a small first generation mass $m_1^2=10^{-6}$ eV$^2$ \cite{Li:2023ncg}, and input 
$m_2$ and $m_3$ from the measured mass-squared differences 
$\Delta m^2_{21} \equiv m^2_{2}-m^2_{1}= (7.50^{+0.22}_{-0.20})\times 10^{-5}$ eV$^2$ 
and $\Delta m^2_{31}\equiv m^2_{3}-m^2_{1}=(2.55^{+0.02}_{-0.03})\times 10^{-3}$ eV$^2$ in the
NO scenario \cite{deSalas:2020pgw}. The values of $u_1$ and $v_1$ are then 
chosen to render $|G_{1r}|$ and $|G_{1i}|$ have similar deviation from zero
around the two roots of $u_1$ from Eq.~(\ref{s1r}). Note that $v_1$ can be determined only
up to a sign as observed in \cite{Li:2023ncg}. We obtain
\begin{eqnarray}
r_1\approx -0.83-0.04i,\;\;\;\;-0.98+0.04i,\label{pr1}
\end{eqnarray}
which are insensitive to the variation of $m_1$. The sign of $v_1$ has been
assigned in the way that the former (latter) solution corresponds to the $\mu e$ 
($\tau e$) mixing. Compared with the ratios $r_1\approx -1.0-0.02i$ ($r_1\approx -1.0+0.02i$) 
for the $\mu e$ ($\tau e$) mixing derived in the three-generation case \cite{Li:2023ncg}, 
the real part increases to about $-0.83$ (remains similar). 

%, i.e., deviate from zero equally
%minimizing the deviation $|G_{1r}|+|G_{1i}|$ from zero under the condition that
%After Eq.~(\ref{ni}) is inserted, $G_{1r}$ depends on $v_1^2$, 
%while $G_{1i}$ is independent of $v_1$ as stated above. Therefore, 

%To keep two minima of the deviation at 
%$u_1\approx -0.83$ and $-0.98$, which will be assigned as the solutions for the $\mu e$ and 
%$\tau e$ mixings separately, $|v_1|$ cannot exceed 0.04, above which the minimum of the 
%deviation appears only at $u_1\approx -0.91$. 
%The imaginary parts should be understood as an order-of-magnitude estimate. 

%Scanning the above expression in the $u_1$-$u_3$ plane, we find two minima around
%$(u_1,u_3)\approx (-0.8,-0.2)$ and $(-1.0,0.0)$ as indicated Fig. The values of $F_r^2$
%between $(10^-24,10^-22)$ are plotted upside down, such that the grey peak area
%represents the minima of $F_r^2$. It is seen that the minima respect the approximate
%unitarity $u_1+u_3\approx -1$. Substituting the above sets of $u_1$ and $u_3$
%into the solutions for $v_1$ and $v_3$, we get $v_1\approx -v_3\approx -0.2$ and $+0.2$,
%respectively. As explained in, it happens that the $\mu e$ mixing picks up the solution
%$r_1\approx -0.8-0.2i$ and the $\tau e$ mixing picks up $r_1\approx -1.0+0.2i$. Our 
%predictions agree well with the data

The mixing angles $\theta_{12}= (34.3\pm 1.0)^{\circ}$, 
$\theta_{13}=(8.53^{+0.13}_{-0.12})^{\circ}$ and $\theta_{23}=(49.26\pm 0.79)^{\circ}$, 
and the $CP$ phase $\delta=(194^{+24}_{-22})^{\circ}$ from \cite{deSalas:2020pgw},
and the set of $\theta_{12}= (33.40^{+0.80}_{-0.82})^{\circ}$, 
$\theta_{13}=(8.59^{+0.13}_{-0.12})^{\circ}$, $\theta_{23}=(42.4^{+1.0}_{-0.9})^{\circ}$ 
and $\delta=(223^{+32}_{-23})^{\circ}$ from \cite{Capozzi:2021fjo} yield the measured 
ratios for the $\mu e$ mixing
\begin{equation}           
\frac{U^*_{\mu 1}U_{e1}}{U^*_{\mu 2}U_{e2}}=\left\{ \begin{array}{c}
    -(0.676^{+0.045}_{-0.005}) - (0.072^{+0.118}_{-0.113})i,\\
    -(0.786^{+0.116}_{-0.046}) - (0.178^{+0.094}_{-0.092})i,\\
   \end{array} \right.\label{dat}
\end{equation} 
respectively, and for the $\tau e$ mixing
\begin{equation}           
\frac{U^*_{\tau 1}U_{e1}}{U^*_{\tau 2}U_{e2}}=\left\{ \begin{array}{c}
    -(1.289^{+0.008}_{-0.068})+(0.079^{+0.115}_{-0.125})i,\\
    -(1.260^{+0.106}_{-0.104})+(0.284^{+0.076}_{-0.134})i,\\
   \end{array} \right.\label{Pb}
\end{equation} 
respectively, where the errors mostly come from the variation of $\delta$. Our predictions 
in Eq.~(\ref{pr1}) match the above data within uncertainties, and the sign selection for 
Eq.~(\ref{pr1}) is affirmed. The splitting of the real parts in the four-generation case,
relative to the degenerate real parts in the three-generation case \cite{Li:2023ncg},
improves the consistency between the theoretical and experimental results notably.

An alternative comparison with the data is made through the mixing angles and $CP$ 
phase inferred by the solutions in Eq.~(\ref{pr1}) for the $\mu e$ and $\tau e$ mixings.
These parameters can be deduced by minimizing the deviation
\begin{eqnarray}
\sum_{\mathcal{L}=\mu,\tau}\left|\frac{U^*_{\mathcal{L} 1}U_{e1}}{U^*_{\mathcal{L} 2}U_{e2}}
-(u_\mathcal{L} +v_\mathcal{L} i)\right|,\label{dvi}
\end{eqnarray}
with roughly equal contributions from the two pieces labeled by $\mathcal{L}=\mu,\tau$
for $u_\mu=-0.83$, $v_\mu=-0.04$,
%$v_\mu\in [-0.04,-0.02]$, 
$u_\tau=-0.98$ and $v_\tau=0.04$.
%$v_\tau\in [0.10,0.13]$.
The reason for considering the minimization is that one cannot find roots of the mixing 
angles and $CP$ phase, which vanish the two pieces simultaneously.
It is observed that the mixing angles $\theta_{12}$ and $\theta_{23}$ are correlated with 
each other, and the deviation in Eq.~(\ref{dvi}) decreases 
with $\theta_{12}$. However, a large $\theta_{12}$ renders the domain for minimal deviation
in the $\theta_{13}$-$\delta$ plane broad, such that no definite values of 
$\theta_{13}$ and $\delta$ can be retrieved. Taking these facts into account,
we set $\theta_{12}=34^\circ$, for which the minimization of Eq.~(\ref{dvi}) generates
\begin{eqnarray}
\theta_{23}\approx 47^\circ,\;\;\;\;\theta_{13}\approx 5^\circ,\;\;\;\;\delta\approx 200^\circ.
\end{eqnarray}
The above outcomes are stable against the variation of $m_1$ and, except $\theta_{13}$,  
do not differ much from the data \cite{deSalas:2020pgw,Capozzi:2021fjo}. The angle 
$\theta_{23}$ is close to the maximal mixing $45^\circ$, as concluded in \cite{Li:2023ncg}, 
and slightly prefers to be in the second octant. We mention that the two recent global fits 
lead to different octants for $\theta_{23}$ ($49.3^\circ$ vs $42.4^\circ$) 
\cite{deSalas:2020pgw,Capozzi:2021fjo}.

%The lower $\theta_{13}$ reflects the smaller imaginary parts in Eq.~(\ref{pr1}). 
%suffer larger theoretical uncertainties from 

The angle $\theta_{13}$, assuming a small value, might be impacted by higher-order QED 
corrections to the box diagrams which are responsible for the 
leptonic-state mixing. Below we roughly estimate these effects on the determination 
of $\theta_{13}$. The QED corrections to the box diagrams in the symmetric phase,
where all particles are massless, do not differentiate intermediate channels. 
Hence, the argument for the disappearance of the mixing phenomenon, based on the 
unitarity of the PMNS matrix, still holds at a high energy. The corrections that 
differentiate various channels in the broken phase arise from the box diagrams 
involving intermediate neutrinos of distinct masses and a photon exchanged between two 
external charged leptons. These diagrams give channel-dependent contributions apparently. 
Since $\nu_1$, $\nu_2$ and $\nu_3$ are much lighter than $\nu_4$, the QED effects associated 
with the former are approximately the same, and differ from the one associated with 
$\nu_4$. The former contribution can be inferred from the result for the box diagram 
with light fermions \cite{Braaten:1982yp}, which is destructive to the leading-order 
one. The latter can be inferred from the heavy fermion expansion for the box 
diagram induced by heavy-light currents \cite{Detmold:2021uru}; the leading piece 
of the expansion in the power of $1/m_4$, with the $\nu_4$ line being shrunk to a point,
represents a vertex correction. This contribution can be 
eliminated by choosing a renormalization scale, i.e., in the on-shell renormalization 
scheme. The subleading terms, suppressed by powers of $1/m_4$, are negligible. That is, 
the QED corrections associated with $\nu_4$ are less important.

Relatively speaking, the considered QED effects raise the imaginary parts
$\Gamma_{i4}$, $i=1$, 2 and 3, ($\Gamma_{44}$) by a factor $1+\alpha$ ($1+2\alpha$)
compared to the others, $\alpha$ being of order of the QED coupling $\alpha_e\sim 10^{-2}$. 
Including those correction factors into Eq.~(\ref{cons}), we find that the right-hand side 
of Eq.~(\ref{cs1}) becomes $+(10\alpha/7) m_3^4/m_W^4$, where terms proportional to 
$m_{1,2}^2/m_{W,4}^2$ have been dropped for simplicity. This modification changes the 
right-hand side of Eq.~(\ref{nr}) into $-\sqrt{10\alpha/7}m_3^2/m_W^4\approx -5\times 10^{-26}$ 
(the minus sign works better than the plus sign for increasing $\theta_{13}$), 
but leaves Eq.~(\ref{ni}) untouched. Repeating the same procedure, we solve for the ratio 
of the PMNS matrix elements
\begin{eqnarray}
r_1\approx -0.82-0.07i,\;\;\;\;-1.12+0.07i,\label{pr2}
\end{eqnarray}
which are more consistent with the data in Eqs.~(\ref{dat}) and (\ref{Pb}), respectively. 
The minimization of the deviation in Eq.~(\ref{dvi}) with $u_\mu=-0.82$, $v_\mu=-0.07$, 
$u_\tau=-1.12$ and $v_\tau=0.07$ then leads to the mixing angles and $CP$ phase
\begin{eqnarray}
\theta_{23}\approx 47^\circ,\;\;\;\;\theta_{13}\approx 6^\circ,\;\;\;\;\delta\approx 205^\circ,
\end{eqnarray}
for $\theta_{12}=34^\circ$. It is seen that $\theta_{13}$ can indeed be enlarged by the QED
corrections along with slight increment of $\delta$.

We then examine the dispersive constraints on the PMNS matrix elements in the IO
scenario for the neutrino masses. Two solutions similar to Eq.~(\ref{pr1}) are acquired
\begin{eqnarray}
r_1\approx -1.032-0.005i,\;\;\;\;-1.014+0.005i,
\end{eqnarray}
with $m_3^2=10^{-6}$ eV$^2$, and $m_1^2$ and $m_2^2$ from 
$\Delta m^2_{21} =  (7.50^{+0.22}_{-0.20})\times 10^{-5}$ eV$^2$ and 
$\Delta m^2_{31}= -(2.45^{+0.02}_{-0.03})\times 10^{-3}$ eV$^2$ \cite{deSalas:2020pgw}.
The results are insensitive to the variation of $m_3$ in this case. It is hard to 
associate either of them with the $\mu e$ mixing. Nevertheless, it is clear that the small 
imaginary parts disagree with the measured ratio for the $\mu e$ mixing in the IO scenario,
\begin{eqnarray}
r_1=-(1.05^{+0.23}_{-0.18})- (0.38^{+0.00}_{-0.05})i,\label{da2}
\end{eqnarray}
corresponding to $\theta_{12}=(34.3\pm 1.0)^{\circ}$, 
$\theta_{13}=(8.58^{+0.12}_{-0.14})^{\circ}$, $\theta_{23}=(49.46^{+0.60}_{-0.97})^{\circ}$ and 
$\delta=(284^{+26}_{-28})^{\circ}$ from \cite{deSalas:2020pgw}. In other words,
the dispersive constraints refute the IO spectrum for the neutrino masses as advocated 
in \cite{Li:2023ncg}.

At last, one may suspect that the existence of the massive fourth generation leptons 
would modify the cosmological constraint on the sum of neutrino masses $\sum m_\nu$. 
The neutrino masses have been assumed to be approximately degenerate in the derivation of 
the above constraint. The case with additional significantly massive neutrinos require 
a different treatment as pointed out in \cite{Mantz:2009rj}. This subject can be addressed 
in a future publication. Besides, one may wonder whether the massive fourth generation 
neutrino could contribute to the effective neutrino masses, such as the effective Majorana 
mass $m_{ee}$, through the mixing with the light neutrinos. 
As explained below Eq.~(\ref{au}), the mixing between the light neutrinos and the
fourth generation one is extremely suppressed by the factor $m_i^2/m_W^2$. Therefore, the 
fourth generation neutrino gives $m_4U_{e4}^2\sim m_4m_i^2/m_W^2\approx
10^{-11}$ eV to $m_{ee}$, where the light neutrino mass is taken as $m_i=1$ eV for the 
estimate. This contribution is completely negligible.

\section{CONCLUSION}

We have continued our efforts to constrain the parameters in the SM and
beyond in dispersive analyses of physical observables. Following our previous determination 
for the masses of the sequential fourth generation quarks in an extension of the SM, we
have predicted the masses of the fourth generation leptons by solving the dispersion 
relations associated with heavy fermion decays. The approach has been elucidated in our
previous works and the application to the present investigation is straightforward. The mass 
$m_4\approx 170$ GeV ($m_L\approx 270$ GeV) of the fourth generation neutral (charged) lepton 
with little theoretical errors was extracted from the study of the $t\to d e^+\nu_4$ 
($L^-\to \nu_1 \bar t d$) decay width with the input of the known top quark mass 
$m_t\approx 173$ GeV. The above results were then validated by means of the 
$L^-\to \nu_4 \bar u d$ decay width. Although this mode suffers higher-order QCD 
corrections attributed to the light quark pair in the final state, the cross-check 
provides a solid support to our formalism and predictions. It has been inspected that the 
fourth generation leptons with the aforementioned masses survive the experimental bounds from 
Higgs boson decays into photon pairs and from the oblique parameters. Briefly speaking, 
the fourth generation charged lepton reduces the SM contribution to the former by
37\%, and, together with the neutral lepton, enhance the oblique parameter $T$ to 0.178.
These effects are still within the current experimental uncertainties.

We discussed the dispersive constraints on the neutrino masses and the PMNS matrix 
elements in the presence of the fourth generation leptons. It turns out that the heavy 
neutrino demands the almost exact unitarity of the $3\times 3$ PMNS matrix, and the 
observed ratios of the PMNS matrix elements, $U^*_{\mu 1}U_{e1}/(U^*_{\mu 2}U_{e2})$ and 
$U^*_{\tau 1}U_{e1}/(U^*_{\tau 2}U_{e2})$, can be better accommodated, as the known 
mass-squared differences $\Delta m^2_{21}$ and $\Delta m^2_{32}$ are input. 
This accommodation marks an improvement of our previous analysis based on three generations 
of neutrinos, which, in some sense, suggests the existence of the fourth generation neutrino. 
The constraints on the above two ratios are equivalent to those on the mixing angles and 
$CP$ phase in the $3\times 3$ PMNS matrix, giving $\theta_{12}\approx 34^\circ$, 
$\theta_{23}\approx 47^\circ$, $\theta_{13}\approx 5^\circ$ and $\delta\approx 200^\circ$ 
approximately, close to the data for the NO scenario (except $\theta_{13}$). The above results 
are insensitive to the variation of the lightest neutrino mass $m_1$. The consistencies of 
our theoretical framework in various aspects encourage the search for such heavy neutral 
and charged leptons at the (high-luminosity) large hadron collider or a muon collider.

\section*{Acknowledgement}

We thank K.F. Chen, Y.T. Chien, Y. Chung, X.G. He, W.S. Hou, C.J. Lin, M. Spinrath and 
M.R. Wu for fruitful discussions. 
This work was supported in part by National Science and Technology Council of the Republic of 
China under Grant No. MOST-110-2112-M-001-026-MY3.

%%%%%%%%%%%%%%%%%%%%%%%%%%%%%%%%%%%%%%%%%%%%%%%%%%%%%%%%%%%%%%%%%%%%%%%%%%%%%%%%%%%%%%%%%%%%%%%%%%%%%%%%%%%%%%%%%%%%%%%%%%%%

\end{document}